%% file: main.tex
\documentclass[acmlarge]{acmart}
\setcopyright{acmlicensed}
\acmJournal{IMWUT}
\acmYear{March2024} \acmVolume{8} \acmNumber{1} \acmArticle{41} \acmMonth{1}\acmDOI{10.1145/3643560}

\usepackage{amsmath}
\usepackage{algorithm}
\usepackage{algorithmicx}
\usepackage{algpseudocode}
\usepackage{graphicx}
\usepackage{lipsum}
\usepackage{enumerate}
\usepackage{enumitem}
\usepackage{hyperref}
\usepackage{multirow, makecell}
\usepackage{nccmath}
\usepackage{ragged2e}
\usepackage{threeparttable} 
\usepackage{url}
\usepackage{xspace}
\usepackage{subcaption}
\usepackage{bbding}

\AtBeginDocument{%
  \providecommand\BibTeX{{%
    \normalfont B\kern-0.5em{\scshape i\kern-0.25em b}\kern-0.8em\TeX}}}

\setcopyright{acmcopyright}
\copyrightyear{2024}
\acmYear{2024}

\newcommand{\ie}{\emph{i.e.},\xspace}

\newcommand{\eg}{\emph{e.g.},\xspace}

\newcommand{\etal}{\emph{et al.}\xspace}

\newcommand\figref[1]{Fig.~\ref{#1}}

\newcommand\tabref[1]{Tab.~\ref{#1}}

\newcommand\secref[1]{Sec.~\ref{#1}}

\newcommand{\fakeparagraph}[1]{\vspace{1mm}\noindent\textbf{#1.}}

\newcommand{\sysname}{{\sf EchoPFL}\xspace}

\ifodd 1

\newcommand\rev[1]{\textcolor{black}{#1}}

\else

\newcommand\rev[1]{#1}

\fi
\definecolor{mypurple}{RGB}{128,0,128}

\algrenewcommand{\algorithmiccomment}[1]{\textcolor{black}{/* #1 */}}

\begin{document}

\title{EchoPFL: Asynchronous Personalized Federated Learning on Mobile Devices with On-Demand Staleness Control}

\author{Xiaochen Li}
\orcid{0000-0003-2653-5786}
\affiliation{%
  \institution{Northwestern Polytechnical University}
  \department{School of Computer Science}
  \city{Xi'an}
  \country{China}
}

\author{Sicong Liu}
\orcid{0000-0003-4402-1260}
\affiliation{%
  \institution{Northwestern Polytechnical University}
  \department{School of Computer Science}
  \city{Xi'an}
  \country{China}
    \thanks{Corresponding author: scliu@nwpu.edu.cn}
}

\author{Zimu Zhou}
\orcid{0000-0002-5457-6967}
\affiliation{%
  \institution{City University of Hong Kong}
  \department{School of Data Science}
  \city{Hong Kong}
  \country{China}
}

\author{Bin Guo}
\orcid{0000-0001-6097-2467}
\affiliation{%
  \institution{Northwestern Polytechnical University}
  \department{School of Computer Science}
  \city{Xi'an}
  \country{China}
}

\author{Yuan Xu}
\orcid{0009-0004-2067-9847}
\affiliation{%
  \institution{Northwestern Polytechnical University}
  \department{School of Computer Science}
  \city{Xi'an}
  \country{China}
}

\author{Zhiwen Yu}
\orcid{0000-0002-9905-3238}
\affiliation{%
  \institution{Harbin Engineering University}
  \city{Harbin}
  \country{China}
}
\affiliation{%
  \institution{Northwestern Polytechnical University}
  \department{School of Computer Science}
  \city{Xi'an}
  \country{China}
}

\begin{abstract}
The rise of mobile devices with abundant sensory data and local computing capabilities has driven the trend of federated learning (FL) on these devices. 
And personalized FL (PFL) emerges to train specific deep models for each mobile device to address data heterogeneity and varying performance preferences.
However, mobile training times vary significantly, resulting in either delay (when waiting for slower devices for aggregation) or accuracy decline (when aggregation proceeds without waiting).
In response, we propose a shift towards asynchronous PFL, where the server aggregates updates as soon as they are available.
Nevertheless, existing asynchronous protocols are unfit for PFL because they are devised for federated training of a single global model. 
They suffer from slow convergence and decreased accuracy when confronted with severe data heterogeneity prevalent in PFL. 
Furthermore, they often exclude slower devices for staleness control, which notably compromises accuracy when these devices possess critical personalized data.
Therefore, we propose \sysname, a coordination mechanism for asynchronous PFL. 
Central to \sysname is to include updates from all mobile devices regardless of their latency. 
To cope with the inevitable staleness from slow devices, \sysname revisits model broadcasting. 
It intelligently converts the unscalable broadcast to \textit{on-demand broadcast}, leveraging the \textit{asymmetrical bandwidth} in wireless networks and the dynamic clustering-based PFL. 
Experiments show that compared to status quo approaches, \sysname achieves a reduction of up to 88.2\% in convergence time, an improvement of up to 46\% in accuracy, and a decrease of 37\% in communication costs.
\end{abstract}



\maketitle

\input{body/introduction.tex}
\input{body/motivation}

\input{body/overview}

\input{body/cluster}
\input{body/broadcast}

\input{body/Implementation}

\input{body/experiment}
\input{body/related}
\begin{acks}
This work was partially supported by the National Science Fund for Distinguished Young Scholars (62025205) the National Natural Science Foundation of China (No. 62032020, 62102317), and CityU APRC grant No. 9610633.
The authors thank Lei Wu for the mathematical discussions about \sysname and the anonymous reviewers for their constructive feedback that has made the work stronger.
\end{acks}

\newpage
\bibliography{acmart}
\bibliographystyle{ACM-Reference-Format}
\end{document}

%% file: body/introduction.tex
\section{Introduction}
\label{sec:intro}
The rapid growth of sensory data generated from ubiquitous mobile devices, coupled with their local computing power, in addition to the widespread availability of wireless networks, has catalyzed the emergence of federated learning (FL) on these devices.
In this paradigm, multiple \textit{clients}, \ie ubiquitous mobile devices such as smartphones, wearables, drones, and robots, collaboratively train a shared model in a specific application scenario under \textit{server} orchestration while keeping their datasets decentralized \cite{mcmahan2017communication, li2021hermes,lai2021oort}. 
FL offers an avenue for the development of data-intensive deep learning applications with ubiquitous mobile devices, including activity recognition \cite{ouyang2021clusterfl, li2023hierarchical}, personalized recommendation \cite{niu2020billion,wang2021fast,liang2021fedrec++}, and transportation ~\cite{liang2022federated, he2022automatch, wang2022fed}. 
Attributed to diverse user behaviors and preferences, sensory data from ubiquitous mobile devices are often non-IID (identically and independently distributed).
For example, one user prefers outdoor activities, while another prefers indoor hobbies. 
Their sensory data, such as GPS location or activity trackers, would exhibit distinct patterns.
This makes it challenging to learn a single model for all clients with high accuracy \cite{tan2022towards, chen2022pfl}. 

To handle such natural \textit{data heterogeneity} on ubiquitous mobile devices, \textit{personalized federated learning (PFL)} has been introduced.
PFL seeks to train client-distinct models to accommodate the diverse data distributions across different clients \cite{tan2022towards}.
PFL strategies roughly fall into \textit{global model personalization} or \textit{learning personalized models}.
The former includes techniques such as local fine-tuning \cite{zhang2022fine, wang2023fedftha, zeng2022gradient} and meta-learning \cite{fallah2020personalized}, while the latter embraces methods like clustering \cite{ouyang2021clusterfl, briggs2020federated, ghosh2020efficient,li2023hierarchical}, multi-task learning \cite{smith2017federated}, and knowledge distillation \cite{zhang2021parameterized}.

However, there exists a significant \textit{gap} when it comes to the practical deployment of such PFL systems in real-world ubiquitous mobile application scenarios.
The gap is the variations in local training time, especially due to the diverse computing resources and network availability of different devices.
Most PFL frameworks, such as those proposed in studies like \cite{ouyang2021clusterfl, briggs2020federated, ghosh2020efficient, smith2017federated, hanzely2020federated, zhang2021parameterized, zhang2023dm-pfl}, are developed with the assumption of \textit{synchronous} model aggregation, where the server waits for updates from all clients in each round.
In essence, \textit{slow devices} can induce significant waiting time and thus the training delay. 
However, collaborative training with mobile devices is always latency-sensitive in ubiquitous applications, for example~\cite{cho2022flame,zhang2022federated}.
Compounding this issue, the specialized designs in PFL often come with even higher computation and communication costs than the non-personalized counterparts \cite{chen2022pfl}.
This brings us to a potential solution: \textit{asynchronous PFL}, where the server aggregates updates when they arrive from clients, eliminating the waiting time associated with stragglers.
In particular, many alternative solutions exist to address the challenge of \textit{mobile system heterogeneity}, such as client selection \cite{nishio2019client, lai2021oort, li2022pyramidfl}, adaptive learning rate control \cite{xu2021learning}, and heterogeneous model architectures \cite{diao2020heterofl,deng2022tailorfl}. 
Among them, the asynchronous strategy is notable, offering simplified client-server coordination and enhanced adaptability to mobile device resource fluctuations~\cite{yang2022anarchic}.

Nevertheless, asynchronous PFL with mobile devices introduces its own set of challenges. 
Asynchronous protocols might incur excessive communication overhead and degraded accuracy due to model \textit{staleness} \cite{xie2019asynchronous, nguyen2022deep, park2021sageflow}.
To balance model accuracy and training latency, semi-asynchronous FL \cite{sun2022fedsea, wu2020safa, ma2021fedsa} \rev{has} been introduced, where clients synchronize with the server at carefully controlled frequencies.
However, we note that these protocols are primarily designed for training a single model and may encounter challenges when applied in the context of PFL (see \secref{sec:pre}).
Also, researchers \cite{yang2022anarchic, dun2023efficient} report slow convergence and considerable accuracy drop of asynchronous protocols with severe data heterogeneity—an issue that is present in PFL.
A more notable concern is the potential exclusion of slower devices for staleness control \cite{wu2020safa}.
When the slow devices contain large amounts of important personalized data, excluding them from training would drastically deteriorate the model accuracy \cite{sun2022fedsea}.

In this paper, we propose \sysname, a simple yet effective client-server coordination mechanism via \textit{proactive on-demand model broadcast} for staleness control in asynchronous PFL with mobile devices. 
At its core, \sysname ensures no critical data is left behind by including updates from all devices, regardless of their local training latency.
Specifically, to manage the inevitable model staleness from slower devices, the server timely broadcasts the most recent aggregated models to clients involved in the training of the same personalized cluster, akin to an "echo".
Note that proactive broadcast is seldom applied in prior asynchronous FL systems because it introduces excessive server-client communication. 
Instead, we find broadcasts suited for asynchronous PFL with mobile devices based on the following \textbf{\textit{novel observations}}: 
\begin{itemize}
    \item Wireless networks connecting ubiquitous devices often feature \textit{asymmetric bandwidth}. 
    For example, the downstream bandwidth can be up to $10 \times$ larger than the upstream bandwidth, in a typical 5G network \cite{chen2014requirements}.
    This \textit{asymmetry} is particularly advantageous because the model broadcast from the server exclusively utilizes the downstream traffic, preventing network congestion and thus long latency.
    \item The data heterogeneity in PFL can be seen as \textit{blessing} rather than a burden with the model broadcasting. 
    This is because data heterogeneity allows for a more targeted approach to broadcasting model updates. 
    By confining model broadcasts to clients with similar data distributions, the scope of the broadcast is constrained.
    Moreover, since broadcast takes place on relatively homogeneous data, it also increases the tolerance to staleness and potentially reduces the broadcast frequency \cite{ouyang2021clusterfl}.
\end{itemize}

Specifically, the design of \sysname is grounded upon clustering-based PFL frameworks \cite{ouyang2021clusterfl, briggs2020federated, ghosh2020efficient}, as many datasets in mobile computing applications exhibit high clusterability \cite{ouyang2021clusterfl}.
To rapidly and accurately assign clients to appropriate clusters based on asynchronously arrived information, \sysname adopts data-aware dynamic client clustering to incrementally create and manage clusters.
Moreover, \sysname periodically merges/expands clusters in line with potential drifts in mobile client data.
Within each cluster, \sysname also predicts the optimal broadcast frequency to further reduce the downstream communication cost without compromising accuracy.
We implement \sysname as a continuous integration (CI) based client-server coordination scheme
which makes it promising for integration with mainstream FL frameworks, such as FLOWER \cite{beutel2020flower}.
We evaluate the performance of \sysname on four mobile tasks and four real-world scenarios with diverse data or system heterogeneity using twenty mobile devices. 
Results show a reduction of up to $88.2 \%$ in training time and up to 37\% in communication cost with an improvement of up to 41.04\% in accuracy.
Especially, \sysname achieves up to $46 \%$ accuracy increase in slow clients
(\secref{sec:experiment}).
Our main contributions are summarized as follows.
\begin{itemize}
    \item 
    To the best of our knowledge, this is the first work that effectively integrates ubiquitous system asynchrony into personalized FL.
    It not only ensures the inclusion of slower mobile clients but also effectively tackles the challenge of model staleness without compromising any mobile model accuracy.
    \item 
    We propose \sysname, asynchronous personalized FL with mobile devices via on-demand model broadcast.
    It harnesses the data heterogeneity in personalized FL and the bandwidth asymmetry in wireless networks via data-aware dynamic client clustering and in-cluster adaptive model broadcast.
    We also implement \sysname as an easy-to-use client-server coordination scheme for integration with other FL frameworks.
    \item 
    Experiments show that \sysname outperforms existing a-/semi-/synchronous or personalized FL methods \cite{mcmahan2017communication, ouyang2021clusterfl, xie2019asynchronous, lai2021oort} in trading off between accuracy and training time at low communication costs across various mobile tasks, platforms, and scenarios.
    It also yields a significant accuracy increase for slow devices.
\end{itemize}

%% file: body/motivation.tex
\section{Motivation and Challenge}
\label{sec:pre}

In this section, we delve into the potential advantages and challenges of asynchronous PFL, drawing insights from preliminary studies on two mobile applications.





\subsection{Motivation for Asynchronous Personalized FL}
\label{sec:pre:motivation}
In the context of mobile applications, FL systems aim to learn \textit{accurate} deep models with \textit{low latency}, allowing for fast adaptation to dynamic contexts and user preferences. 
Generally, Federated Learning (FL) constitutes a collaborative training process between multiple mobile clients and a server \cite{mcmahan2017communication, li2021hermes, lai2021oort, wang2023distribution}.  
Mobile clients perform local training with their local datasets and subsequently upload their trained deep models to the server. 
The server aggregates these models, often using weighted averaging, and then distributes the updated model back to the mobile clients. 
This iterative process continues until convergence.

However, standard FL algorithms, such as FedAvg \cite{mcmahan2017communication}, encounter performance degradation when applied to mobile devices.
The reasons are two-fold:
\begin{itemize}
    \item 
    A unified global model yields low accuracy with non-IID sensory data across mobile devices, necessitating various \textit{personalized} global models for individual mobile devices \cite{tan2022towards, chen2022pfl}. 
    \item 
    The primary \textit{latency bottleneck} in each training round is the need to wait for stragglers, \eg low-speed computing devices.
    This highlights the importance of considering \textit{asynchrony} from a system perspective \cite{xie2019asynchronous, nguyen2022deep, park2021sageflow}.
\end{itemize}

\begin{table}[t]
\tiny
\caption{Performance comparisons of different FL paradigms on two widely used mobile applications.}
\begin{tabular}{|c|cccc|cccc|}
\hline
\multirow{2}{*}{\textbf{Representative paradigms}} & \multicolumn{4}{c|}{\textbf{Image recognition (IR)}} & \multicolumn{4}{c|}{\textbf{Human activity recognition (HAR)}} \\ \cline{2-9} 
 & \multicolumn{1}{c|}{\textbf{\begin{tabular}[c]{@{}c@{}}Accuracy \\ (\%)\end{tabular}}} & \multicolumn{1}{c|}{\textbf{\begin{tabular}[c]{@{}c@{}}Accuracy at the\\ slowest device (\%)\end{tabular}}} & \multicolumn{1}{c|}{\textbf{\begin{tabular}[c]{@{}c@{}}Accuracy at the\\ fastest device (\%)\end{tabular}}} & \textbf{\begin{tabular}[c]{@{}c@{}}Latency \\ (min)\end{tabular}} & \multicolumn{1}{c|}{\textbf{\begin{tabular}[c]{@{}c@{}}Accuracy\\ (\%)\end{tabular}}} & \multicolumn{1}{c|}{\textbf{\begin{tabular}[c]{@{}c@{}}Accuracy at the\\ slowest device (\%)\end{tabular}}} & \multicolumn{1}{c|}{\textbf{\begin{tabular}[c]{@{}c@{}}Accuracy at the\\ fastest device (\%)\end{tabular}}} & \textbf{\begin{tabular}[c]{@{}c@{}}Latency\\ (min)\end{tabular}} \\ \hline
\textbf{Sync FL (FedAvg \cite{mcmahan2017communication})} & \multicolumn{1}{c|}{52.1 $\pm$ 15.3} & \multicolumn{1}{c|}{30.5} & \multicolumn{1}{c|}{60.6} & 398 & \multicolumn{1}{c|}{89.7$\pm$5.3} & \multicolumn{1}{c|}{82.1} & \multicolumn{1}{c|}{94.6} & 37.2 \\ \hline
\textbf{Async FL (FedAsyn \cite{xie2019asynchronous})} & \multicolumn{1}{c|}{48.6 $\pm$ 22.5} & \multicolumn{1}{c|}{22.1} & \multicolumn{1}{c|}{71.3} & 102 & \multicolumn{1}{c|}{85.6$\pm$11.4} & \multicolumn{1}{c|}{65.4} & \multicolumn{1}{c|}{91.2} & 17.4\\ \hline
\textbf{Sync PFL (ClusterFL \cite{ouyang2021clusterfl})} & \multicolumn{1}{c|}{92.4 $\pm$ 3.7} & \multicolumn{1}{c|}{89.6} & \multicolumn{1}{c|}{94.3} & 321 & \multicolumn{1}{c|}{99.6$\pm$0.2} & \multicolumn{1}{c|}{98.9} & \multicolumn{1}{c|}{100} &  33.6\\ \hline
\end{tabular}
\label{tg_motivate_case}
\end{table}

\begin{figure*}[t]
  \centering
  \subfloat[Sync FL]{
    \includegraphics[height=0.145\textwidth]{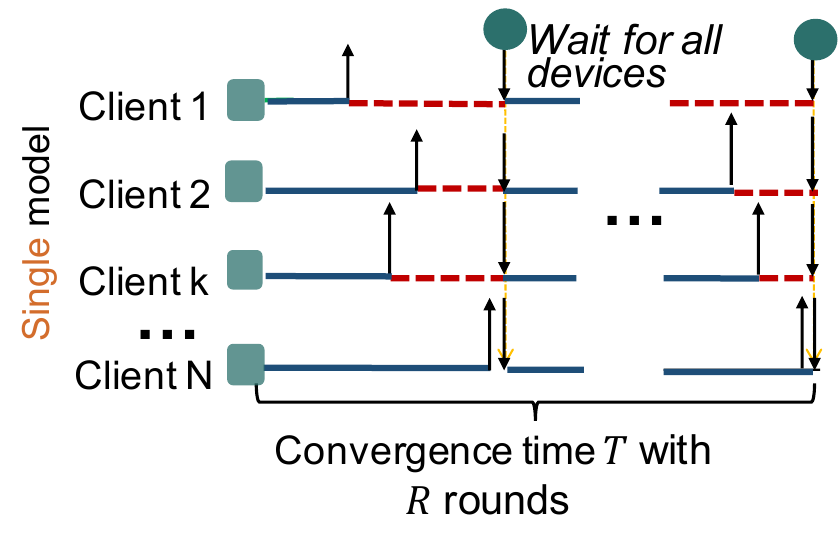}
    \label{fig:over_syn}
  }
  \subfloat[Async FL]{
    \includegraphics[height=0.145\textwidth]{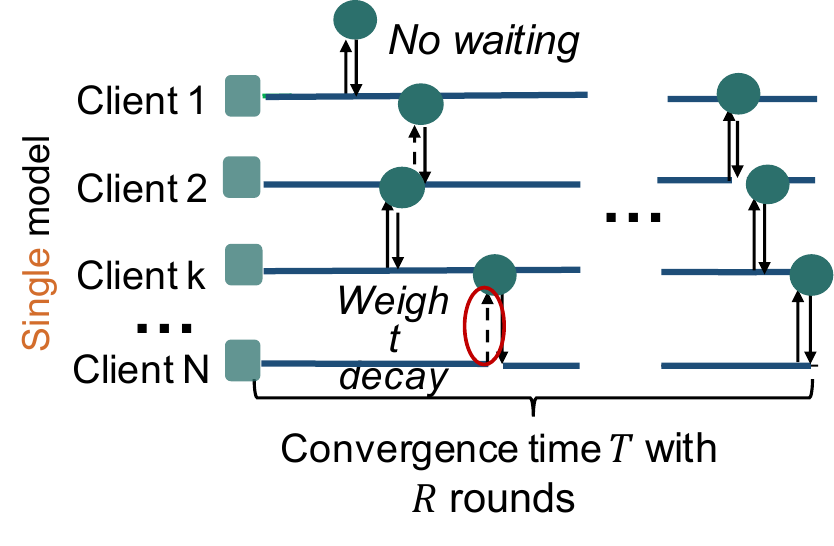}
    \label{fig:over_asyn}
  }
  \subfloat[Sync PFL]{
    \includegraphics[height=0.145\textwidth]{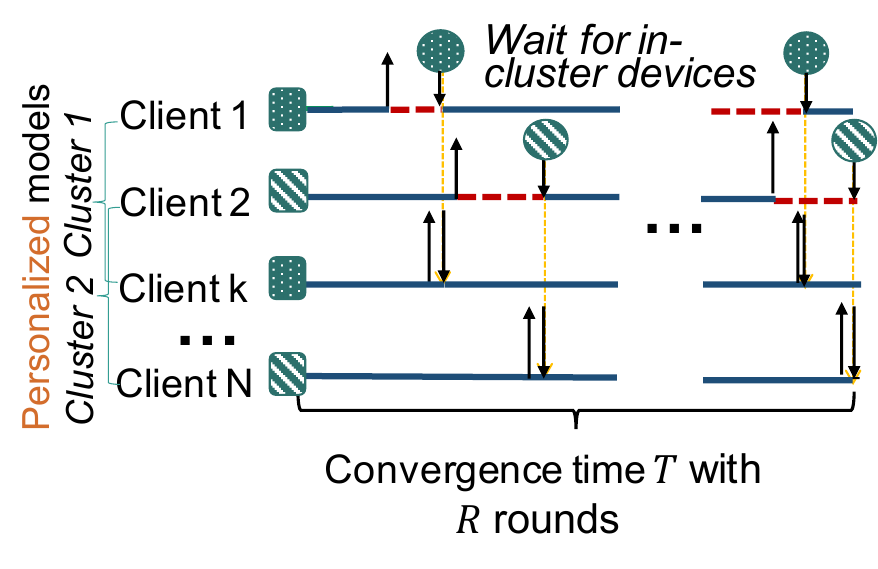}
    \label{fig:over_p_syn}
  }
  \subfloat[Async PFL]{
    \includegraphics[height=0.145\textwidth]{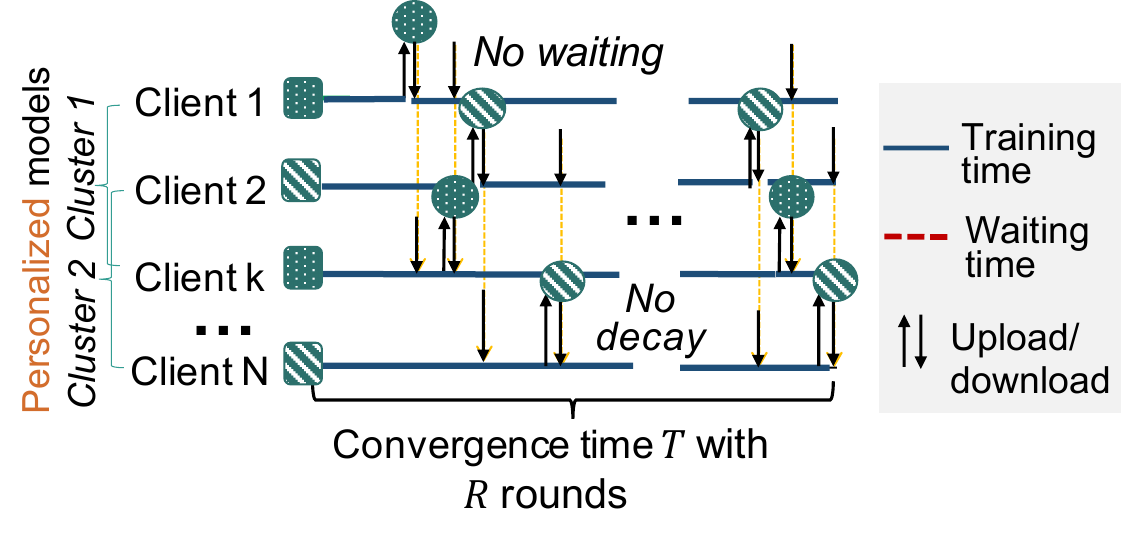}
    \label{fig:over_p_asyn}
  }
  \vspace{-10pt}
  \caption{Potential gains in training latency of asynchronous PFL over existing FL paradigms.}
  \vspace{-10pt}
  \label{fig:async-sync compare}
\end{figure*}

Current FL paradigms with mobile devices that emphasize either \textit{data} or \textit{resource} heterogeneity  fail to strike an optimal balance between model accuracy and training latency.
To illustrate, we evaluate the performance of representative paradigms with two tasks: \textit{image recognition} and \textit{human activity recognition}. 
The former encompasses applications such as smartphone-based face recognition ~\cite{kim2022adaface} and robot-driven security patrolling ~\cite{do2021smart}. 
The latter is widely used in health-care~\cite{straczkiewicz2021systematic,zhang2022quantifying,jung2022imu,wu2020emo} and motion detection \cite{zhuang2019sport}. 
\tabref{tg_motivate_case} presents the results on the CIFAR-10 \cite{krizhevsky2009learning} (for image classification) and HAR-UCI  \cite{anguita2013public} (for human activity recognition) employing 12 diverse mobile devices using three FL paradigms: synchronous FL (FedAvg \cite{mcmahan2017communication}), asynchronous FL (FedAsyn \cite{xie2019asynchronous}), and synchronous PFL (ClusterFL \cite{ouyang2021clusterfl}). 
The detailed experimental setups are available in \secref{sec:experiment:setup}. 
Note that our focus on clustering-based PFL algorithms arises from the high clusterability observed in the data distributions across many mobile applications \cite{cao2018gchar, stisen2015smart, ouyang2021clusterfl}.
We make the following observations.
\begin{itemize}
    \item 
    Synchronous personalized FL (PFL) significantly outperforms synchronous FL in \textbf{\textit{accuracy}}. 
    In image recognition, synchronous PFL improves the accuracy by at least 40.3\% over synchronous FL. 
    For human activity recognition, it shows a 9.9\% accuracy advantage over synchronous FL.
    \item 
    The \textbf{\textit{accuracy}} of synchronous personalized FL is also 4.1\% higher than asynchronous FL (16.7\% higher on the lowest device), yet with longer training latency.
    \item 
    Asynchronous FL drastically decreases training \textbf{\textit{latency}}. 
    In image recognition, it reduces the training time by 74.4\% compared to synchronous FL, whereas in human activity recognition, the reduction is 53.2\%.
\end{itemize}

These observations motivate us to not only adopt the PFL paradigm to ensure the accuracy of different ubiquitous mobile applications but also to further reduce the training latency of PFL by transitioning from synchronous to asynchronous model aggregation, \ie asynchronous PFL.
The intuition is also demonstrated in \figref{fig:async-sync compare}. 
Specifically, asynchronous FL effectively reduces latency by eliminating the waiting time for slow devices during model aggregation (\figref{fig:over_asyn} vs. \figref{fig:over_syn}).
Synchronous PFL (via clustering) employs personalized models to balance generalization and personalization, meanwhile accelerates the training process by only waiting for the slow devices within the same clusters (\figref{fig:over_p_syn} vs. \figref{fig:over_syn}).
Notably, the training latency of synchronous PFL can be further optimized by enabling asynchronous aggregation within clusters (\figref{fig:over_p_asyn} vs. \figref{fig:over_p_syn}).
Therefore, by shifting to asynchronous PFL, we would achieve both higher model accuracy and lower training latency.

\subsection{Challenges in Asynchronous PFL with Mobile Devices}
\label{sec:pre:challenge}
While asynchronous protocols \cite{xie2019asynchronous, nguyen2022deep, park2021sageflow} and personalization techniques \cite{ouyang2021clusterfl, briggs2020federated, ghosh2020efficient, smith2017federated, hanzely2020federated, zhang2021parameterized} exist for FL, seamlessly integrating the two introduces non-trivial challenges.

\begin{figure}
  \centering
  \subfloat[Synchronous PFL]{
    \includegraphics[height=0.16\textwidth]{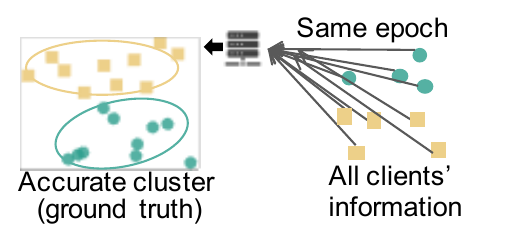}
    \label{fig:distance(a)}
  }
  \hspace{15pt}
  \subfloat[Asynchronous PFL]{
    \includegraphics[height=0.16\textwidth]{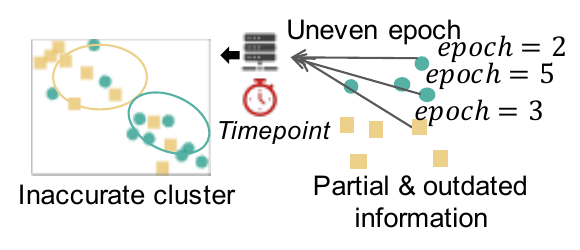}
    \label{fig:distance(b)}
  } 
  \vspace{-10pt}
  \caption{Impact of asynchrony on clustering.}
  \vspace{-10pt}
  \vspace{-3mm}
  \label{fig:distance_comp}
\end{figure}

\textbf{Challenge \#1: How to adapt the personalization strategies to the asynchronous mobile system setting?}
Prior PFL frameworks \cite{ouyang2021clusterfl,sattler2020clustered} implicitly assume a synchronous, globally comprehensive information collection process for subsequent processes, \eg clustering.
However, in real-world asynchronous system settings, where mobile clients may not respond to the server within the desired time limit, the server must cluster mobile clients based on partial information. 
This process must also operate under real-time constraints, potentially leading to less accurate clustering results.
For example, as shown in \figref{fig:distance_comp}, existing synchronous PFL methods cluster mobile clients \rev{by}
utilizing the same epoch to measure specific probabilistic distance metrics among the local model updates returned by all mobile clients~\cite{ouyang2021clusterfl, sattler2020clustered}.
However, when local model updates inevitably arrive at the server in different epochs, this approach fails to promptly and accurately identify clusters.

\begin{figure}[t]
  \centering
  \includegraphics[width=0.48\textwidth]{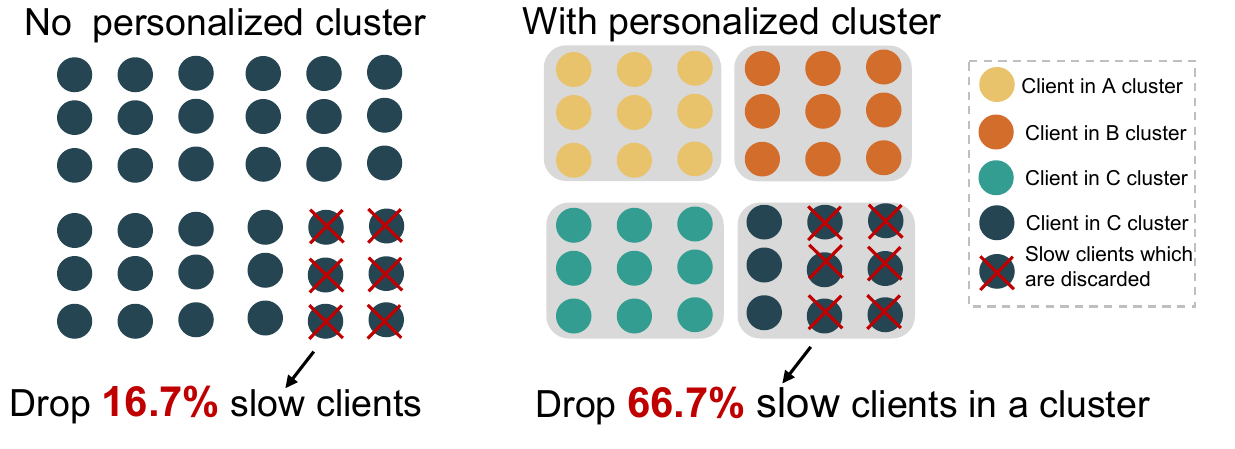}
  \caption{Impact of slow device model dropping on PFL.}
  \vspace{-3mm}
  \label{fig:drop}
\end{figure}

\textbf{Challenge \#2: How to control the model staleness without decaying or discarding updates from slow mobile devices in personalized FL process?}
Asynchronous FL protocols often adopt weight decay \cite{xie2019asynchronous,park2021sageflow} or model dropping \cite{wu2020safa, ma2021fedsa} to mitigate the negative impact on aggregated model accuracy caused by outdated models.
However, we argue that these methods can result in a significant degradation in the accuracy of personalized models. 
This is because slow devices may possess crucial data necessary for specific personalized models. 
To illustrate this point, consider the toy example presented in Figure \ref{fig:drop}.
Excluding 6 slow devices from a conventional FL (\eg FedAvg \cite{mcmahan2017communication}) neglects a mere 16.7\% of data for the single model (marked in blue). 
Conversely, discarding the same 6 devices in clustering-based PFL (\eg ClusterFL \cite{ouyang2021clusterfl}) leads to a pronounced data loss of 66.7\% for that cluster (marked in blue). 
Such data omission can significantly damage the model accuracy for that cluster.

%% file: body/overview.tex
\section{Solution Overview}
\label{sec:overview}

This section presents an overview of \sysname, a mobile client-server coordination mechanism for asynchronous PFL.
\sysname resolves \textit{Challenge \#1} in \secref{sec:cluster} through \textit{dynamic client clustering}, which remains efficient and effective despite the asynchronous arrival of mobile client information.
\sysname addresses \textit{Challenge \#2} in \secref{sec:broadcast} by revisiting \textit{model broadcast} mechanism, previously deemed unscalable for model staleness control in FL, and refining it into \textit{on-demand model broadcast}.

\begin{figure*}[t]
  \centering
  \includegraphics[width=0.95\textwidth]{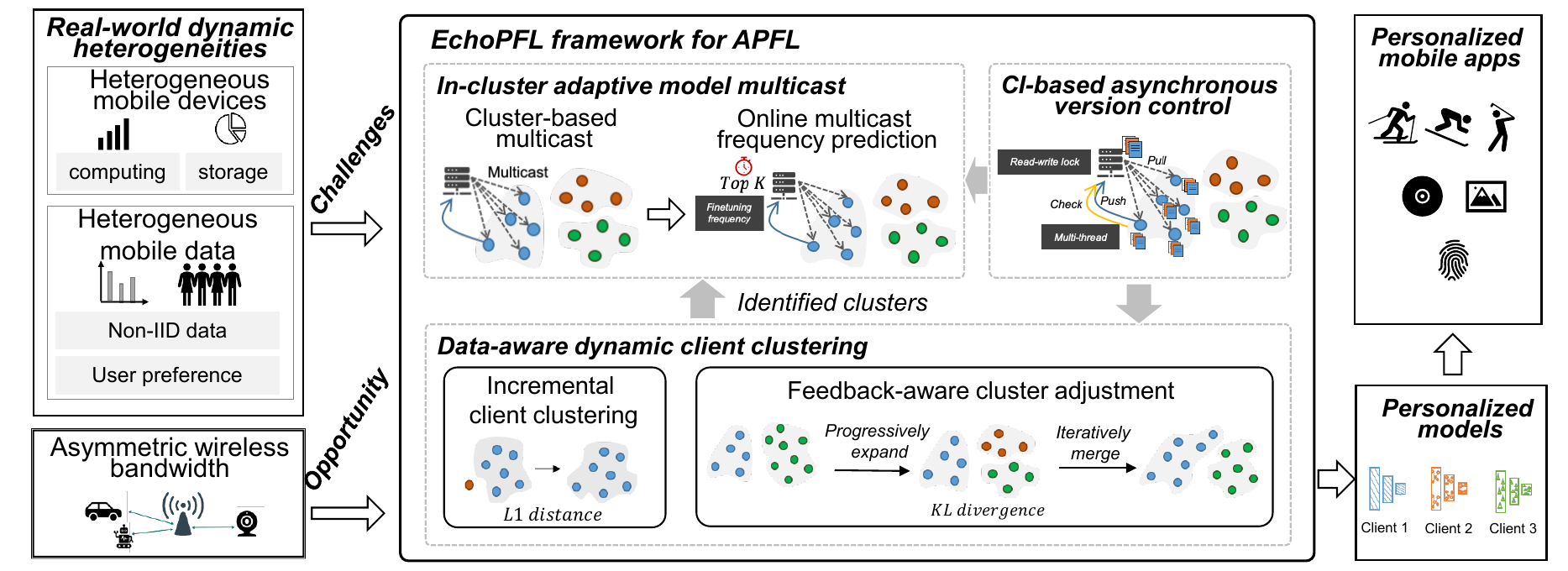}
  \vspace{-10pt}
  \caption{\rev{Overview of \sysname for asynchronous PFL with mobile devices.}}
  \label{fig:overview}
  \vspace{-10pt}
\end{figure*}
\fakeparagraph{Design Rationales}
As mentioned in \secref{sec:intro}, \textit{on-demand model broadcast} becomes viable for model staleness control in PFL with mobile devices, given the inherently limited model broadcast scope in clustering-based PFL.
The core principle guiding the management of model staleness in PFL is the imperative need to ensure that the model updates used in the training process remain relatively up-to-date. 
This necessity becomes particularly critical in light of the asynchronous nature of ubiquitous mobile systems, which can introduce delays in delivering specific mobile model updates to the server. 
These delays may potentially lead to the utilization of outdated models during the aggregation process and subsequent training rounds.
In essence, on-demand model broadcast addresses these challenges by enabling the broader and more timely dissemination of model updates. 
This is a departure from the traditional asynchronous FL, where updates are typically confined to specific devices. 
The broader broadcast scope ensures the rapid distribution of model updates to a wider audience, thereby reducing model staleness.
Consequently, it effectively transforms into an "on-demand model broadcast" problem, which is exclusive to clients sharing similar data distributions, such as those within clusters. 
This approach effectively addresses model staleness while minimizing communication overhead. 
It's worth noting that broadcast operations over homogeneous data, \ie within a cluster, can enhance the system's resilience to staleness, potentially reducing the need for frequent broadcasts~\cite{ouyang2021clusterfl}.
Furthermore, on-demand broadcast leverages the more abundant downstream traffic rather than its upstream link. 
Accordingly, on-demand broadcast from the server to clients is unlikely to trigger notable network congestion and thus latency surges.

\fakeparagraph{System Workflow}
\sysname consolidates the above rationales into an \textit{on-demand model broadcast} mechanism, exemplified within clustering-based PFL frameworks \cite{ouyang2021clusterfl,sattler2020clustered}.
\figref{fig:overview} shows the architecture of \sysname.
It mainly consists of two functional modules:
\begin{itemize}
    \item \textit{Data-aware Dynamic Client Clustering} (\secref{sec:cluster}).
    As client clustering identifies the scope of model broadcast for staleness control, it is crucial to conduct rapid and accurate client clustering in light of asynchronously arriving local model updates. 
    \sysname dynamically creates and manages clusters via on-arrival initial clustering and periodic feedback-aware cluster refinement.
    \item \textit{In-cluster Adaptive Model broadcast} (\secref{sec:broadcast}).
    After clustering, \sysname broadcasts the latest aggregated models to in-cluster clients as required.
    Due to the relatively homogeneous data distribution within a cluster, training is resilient against a certain degree of staleness.
    Also, \sysname predicts the optimal broadcast frequency to further reduce the downstream communication cost without compromising accuracy.
\end{itemize}
To manage asynchronous version conflicts on server and clients, we further implement \sysname as a continuous integration (CI) based client-server coordination scheme for easy integration with other FL frameworks (\secref{sec:implementation}).

%% file: body/cluster.tex
\section{Data-aware Dynamic Clustering}
\label{sec:cluster}
As mentioned in \secref{sec:overview}, rapid and accurate mobile client clustering in light of asynchronously arriving local model updates not only constitutes a critical stage in realizing PFL but also plays an essential role in defining the broadcast scope.
We first discuss the requirements and challenges for client clustering in asynchronous PFL, before explaining the designs of our data-aware dynamic clustering scheme.

\subsection{Primer on Mobile Client Clustering}
\label{sec:cluster_challenges}
Clustering in personalized FL aims to gather mobile clients whose local sensory datasets share similar distributions into the same cluster for training \cite{ouyang2021clusterfl,sattler2020clustered,jothimurugesan2023federated}.
Since the local datasets are inaccessible in FL, the server typically measures the similarity between the models uploaded by the mobile clients as proxy to the similarity of local data distributions.
The similarity between models can be measured at the \textit{parameter} level metric (\eg directly compare model parameters via L1 distance) or at the \textit{feature} level (\eg compare feature maps of model outputs via KL divergence).
Feature-level metrics better reflect the similarity of data distributions among mobile clients than parameter-level metrics \cite{ouyang2021clusterfl}.
Yet the feature-level metrics suffer from longer latency because it needs to collect outputs of model inference.
For \textit{fast} and \textit{accurate} client clustering, \sysname adopts a two-phase scheme by first promptly assigning mobile clients to an initial cluster upon arrival (\secref{sec:cluster_real_time}), and then periodically refining the clusters via feedback from mobile clients (\secref{sec:cluster_adjustment}). 
We elaborate on the two designs below.

\subsection{On-arrival Initial Cluster Assignment of Mobile Devices}
\label{sec:cluster_real_time}
This module immediately assigns a client to a cluster upon receiving its model update to achieve real-time client clustering in the asynchronous setting.

\subsubsection{Mobile Cluster Initialization}
\label{sec:cluster_init}
Since model updates arrive asynchronously, \sysname initializes the clusters incrementally \cite{cao2006density}. 
Given a predefined number of clusters $C$, \sysname initializes the centers of $C$ clusters as the first $C$ local model parameters that arrive at the server.

\subsubsection{Mobile Cluster Assignment}
For the newly arrived model parameters $u_{i}$ from a client $i$, \sysname calculates the L1 distance $L$ between $u_{i}$ to all $C$ clusters and assigns client $i$ to the cluster with the smallest L1 distance:
\begin{equation}
\begin{split}
    &cluster = \mathrm{arg\,min}  \quad (L(u_{i},v_{1}),L(u_{i},v_{2}),\dots,L(u_{i},v_{C}))
    \\&where \quad  L(u_{i},v_{c}) = \|u_{i}-v_{c}\|
\end{split}
    \label{equ:l1_distance}
\end{equation}
$u_{i}$ is the model parameters uploaded by client $i$, and $v_{c}$ represents the model parameters of the $c$-th cluster center. 

We adopt parameter-level similarity metrics rather than feature-level for real-time mobile client clustering.
As we will show in \secref{sec:benchmark}, the feature-level similarity metrics could be $12,000 \times$ slower than the parameter-level counterpart in this process.
Furthermore, feature-level similarity metrics using KL divergence \cite{ouyang2021clusterfl, claici2020model} may incur considerable I/O contention.
Despite its high efficiency, the initial clustering could be erroneous, which results in the feedback-aware clustering refinement, as we will discuss next.

\subsection{Feedback-aware Mobile Cluster Refinement}
\label{sec:cluster_adjustment}
This module improves the accuracy of the initial clusters by periodical refinement, \eg merging and expanding clusters based on the mobile \textit{client feedback}.

\subsubsection{Assessing Clustering Accuracy via Mobile Client Feedback}
We devise a client feedback scheme to assess the clustering accuracy.
The feedback $g(v_{c}, \Pi_{i})$ of client $i$ measures how the model parameters $v_{c}$ from its assigned cluster $c$ fits its local dataset $\Pi_{i}$.
Specifically, client $i$ performs inference using $v_{c}$ on $\Pi_{i}$ and records the distribution $F_{c}$ of the predicted labels.
It then compares $F_{c}$ with the distribution $F_{i}$ of the actual labels and measures their difference via the chi-squared test $\mathcal{X}^2()$.
That is, the feedback of mobile client $i$ is calculated as:
\begin{equation}
\begin{split}
    g(v_{c}, \Pi_{i}) = \mathcal{X}^2(F_{c}, F_{i}) = \sum_{j=1}^{J} \frac{(F_{c}^{j}-F_{i}^{j})^2}{F_{i}^{j}}
\end{split}
\label{equ:idel_chi}
\end{equation}
where $F_{c}^{j}$ is the observed frequency of the $j$-th class, $F_{i}^{j}$ is the expected frequency of the $j$-th class, and $J$ is the number of classes.
As a separate note, we employ the chi-squared test here to assess distribution differences and account for their non-IID nature and discreteness~\cite{bhatia2020midas}.

\begin{figure}[t]
  \centering
    \subfloat[$F_{c}$, $round=1$]{
    \includegraphics[width=0.23\textwidth]{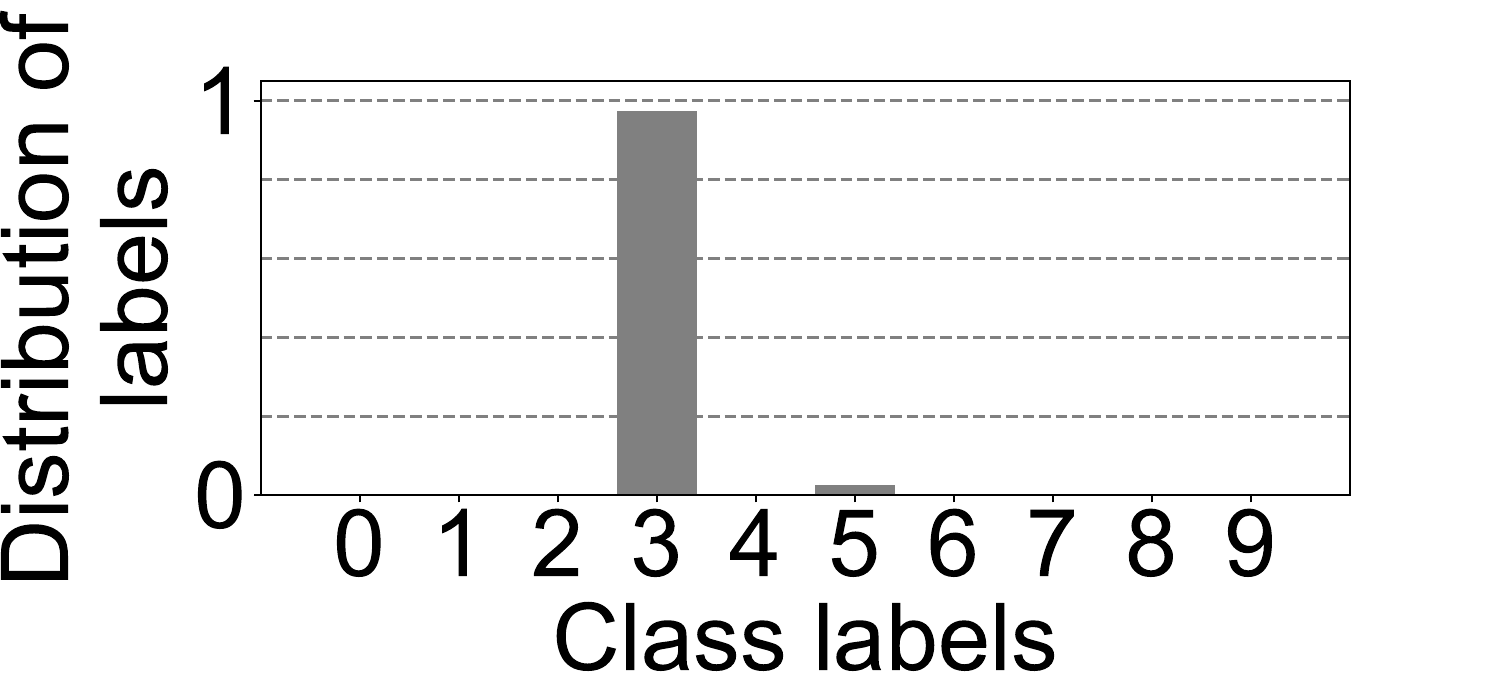}
    \label{fig:error(a)}
  }
  \hfill
  \subfloat[$F_{c}$, $round=100$]{
    \includegraphics[width=0.23\textwidth]{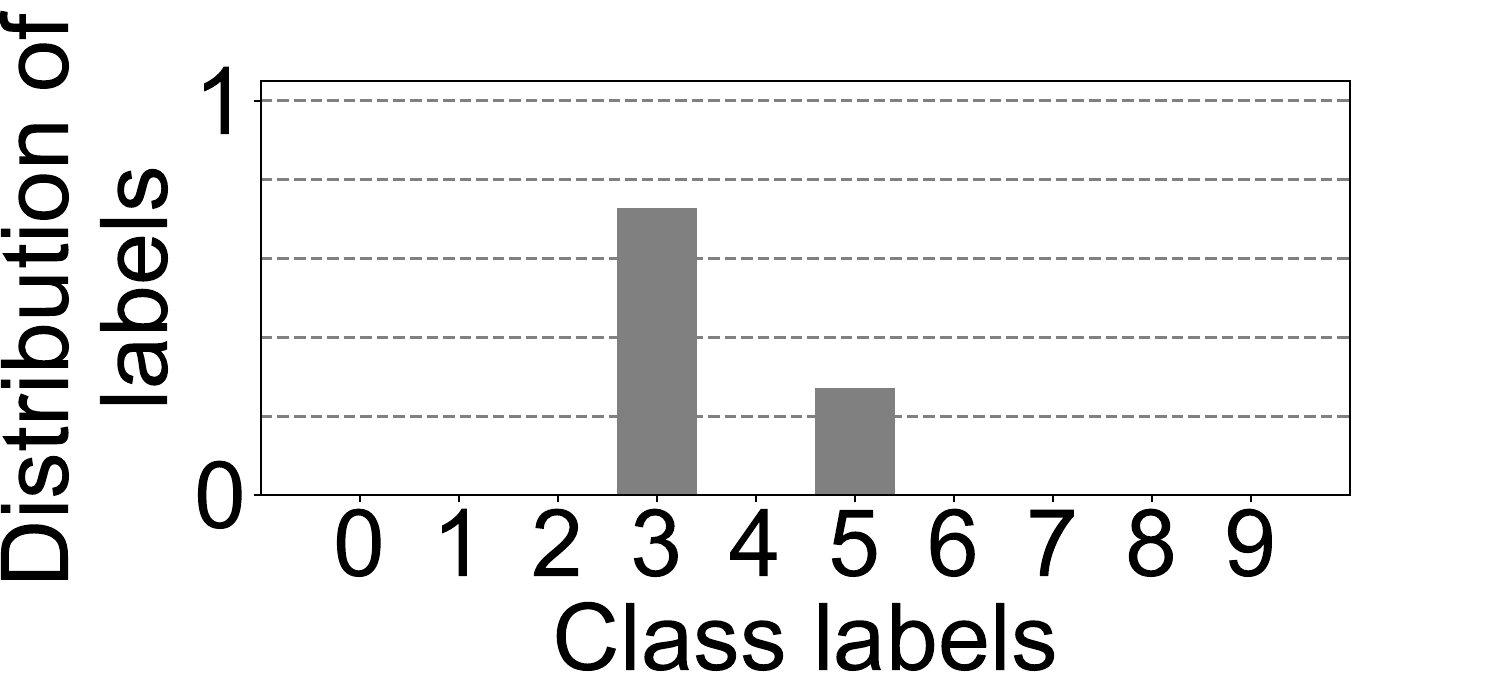}
    \label{fig:error(b)}
  }  
    \hfill
  \subfloat[$S_{c}$, $round=1$]{
    \includegraphics[width=0.23\textwidth]{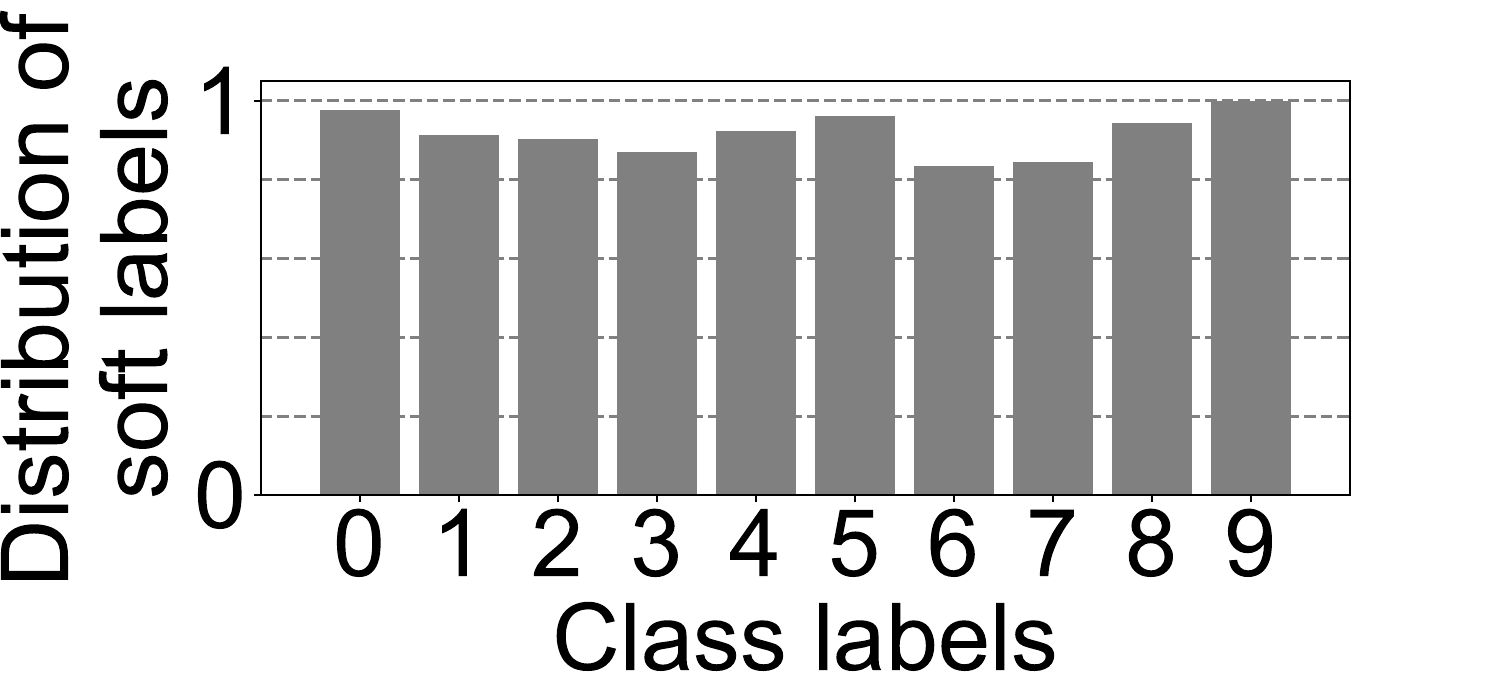}
    \label{fig:error(c)}
  }
  \hfill
  \subfloat[$S_{c}$, $round=100$]{
    \includegraphics[width=0.23\textwidth]{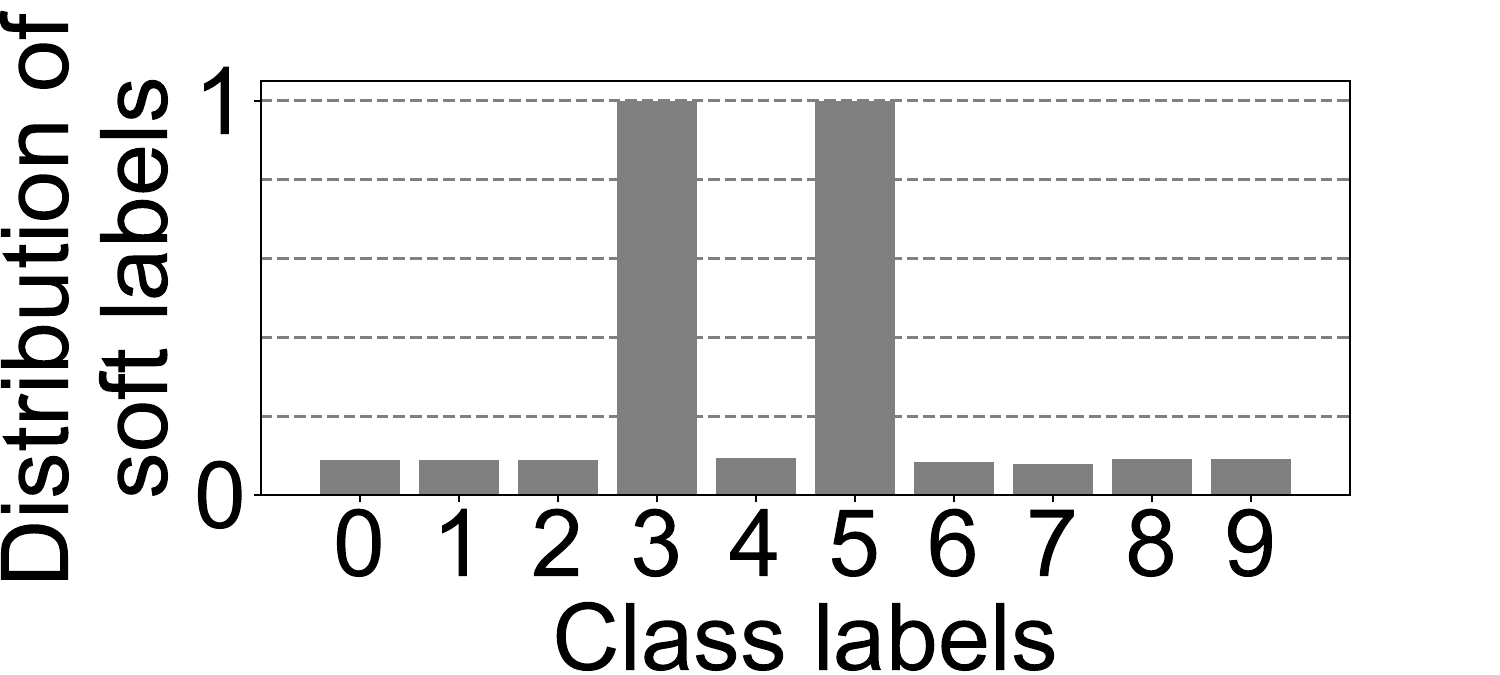}
    \label{fig:error(d)}
  }
  \caption{Illustration of predicted label distribution ($F_{c}$) and soft label distribution ($S_{c}$) based on mobile-side model weights at different training rounds (\ie $round=1$ and $100$) on the same sensory data distribution.}
  \label{fig:error}
\end{figure}

However, the client feedback calculated by $\mathcal{X}^2(F_{c}, F_{i})$ reflects not only the clustering accuracy but also the PFL training stage.
Accordingly, we need to separate and remove the impact of PFL training stages, so that the client feedback only precisely indicates the accuracy of clustering.
\rev{This is non-trivial because the early- and late-stage trained models from slow and fast devices will disturb each other.}
We observe that the distribution of predicted labels $F_{c}$ by $c-th$ cluster center model exhibits distinct patterns during different training stages, eliminating the impact of training from the client feedback.
\figref{fig:error} shows the predicted label distribution $F_{c}$ and the predicted \textit{soft label} distribution $S_{c}$ at different training stages ($round =1$ or $100$).
The soft labels are class probabilities produced by the $c-th$ cluster center model's weights.
It reveals two observations:
\begin{itemize}
    \item The consistency in predicted label distributions shows the viability of employing the chi-squared test to assess label distributions, as highlighted in \figref{fig:error(a)} and \figref{fig:error(b)}.
    \item The predicted distribution of soft labels at various training stages vary, as shown \figref{fig:error(c)} and \figref{fig:error(d)}. 
\end{itemize}
In summary, the model parameters $v_{c}$ from the $c-th$ cluster manifest more significant variances in soft label distributions for predictions across different categories. 
Accordingly, it could be used as a proxy for the training stage, and thus rectify the errors in client feedback.
Specifically, we introduce a probabilistic variance as a measure of \textit{training sufficiency}, and revise the client feedback calculation as follows:
\begin{equation}
\begin{split}
    g(v_{c}, \Pi_{i})   \approx  \sum_{j=1}^{J} \frac{(F_{c}^{j} - F_{i}^{j})^2}{F_{i}^{j}} \cdot \mathrm{Var}(S_{c})
\end{split}
    \label{equ:chi2-fix}
\end{equation}
where $Var(S_{c})$ represents the variance of the predicted distribution of soft labels.

\begin{figure}
    \centering
    
    \subfloat[Mobile cluster merging]{
    \includegraphics[height=0.17\textwidth]{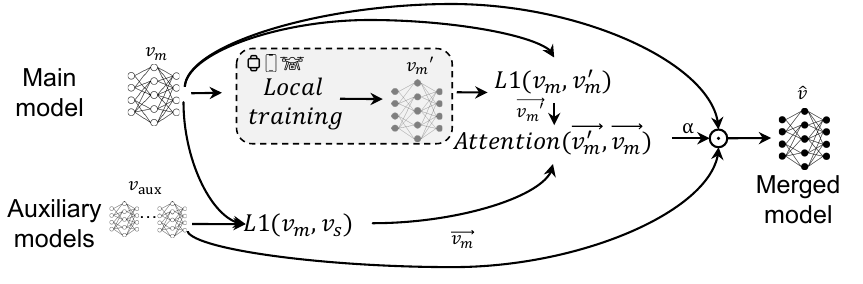}
    \label{fig:expand_merge(a)}
    \vspace{-10pt}
    }
    \subfloat[Mobile cluster expansion]{
    \includegraphics[height=0.17\textwidth]{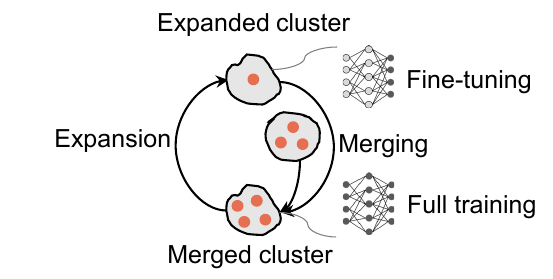}
    \vspace{-10pt}
    \label{fig:expand_merge(b)}
    }
    \vspace{-10pt}
    \caption{Illustration of mobile cluster expansion and merging.}
    \vspace{-10pt}
    \label{fig:expand_merge}
\end{figure}
\begin{algorithm}[t]
\small
\caption{\rev{Cluster merging via optimization direction prediction}}
	\renewcommand{\algorithmicrequire}{\textbf{Input:}}
	\renewcommand{\algorithmicensure}{\textbf{Output:}}
        \label{alg:merging}
    \begin{algorithmic}[1]
        \Require \rev{Cluster center models $v$, local dataset on $i$-th mobile client $\Phi_{i}$;}
        \Ensure \rev{Merged cluster center model $\Hat{v}$;}
        \State \rev{$v_{aux}$(auxiliary models), $v_{m}$(main model) $\leftarrow$ divide $v$;}
        \State \rev{$\vec{v_{m}}$  $\leftarrow$ $L1(v_{aux},v_{m})$ \Comment{\rev{Formulate the assumption for the optimization direction as $\vec{v_{m}}$.};}}
        \State \rev{$v_{m}\textquotesingle$  $\leftarrow$ $F(v_{m},\Phi_{i})$; 
        \Comment{\rev{Train $v_{m}$ on the local dataset $\Phi_{i}$}}}
        \State\rev{$\vec{v_{m}\textquotesingle}$  $\leftarrow$ $L1(v_{m}\textquotesingle,v_{m})$;}\Comment{\rev{Generate posterior distribution $\vec{v_{m}\textquotesingle}$ for optimization direction} }
        \State  \rev{$\alpha$  $\leftarrow $  $Max\{ \vec{v_{m}} \odot \vec{v_{m}\textquotesingle}, 0\}\odot Max\{\vec{v_{m}} \odot \vec{v_{m}\textquotesingle}\}^{-1}$;}\Comment{\rev{Utilize $\vec{v_{m}\textquotesingle}$ and $\vec{v_{m}}$ to generate an aggregate attention map $\alpha$.}}
        \State \rev{$\Hat{v}$ $\leftarrow$ $\alpha\odot v_{aux}+(1-\alpha)\odot v_{m}$} \Comment{ \rev{Employ $\alpha$ to aggregate $v_{m}$ and $v_{aux}$.}}
    \end{algorithmic}
\end{algorithm}

\subsubsection{Mobile Cluster Merging}
\label{subsec:cluster_merging}

As the model parameters of different cluster centers are trained separately for an extended duration, they fit diverse and potentially conflicting local data distributions. 
Naive aggregation of their model parameters as the center of the merged cluster would be sub-optimal~\cite{zhang2022fine,jin2022personalized}.
%
%
To mitigate this, knowledge distillation can avoid direct manipulation of weight parameters.
However, conducting the distillation process typically involves a training phase. 
If it is done on the client side, it will introduce training latency for slow devices and increase memory usage. 
Alternatively, if distillation is performed on the server side, it may compromise asynchronous efficiency due to concurrent distillation processes.

To this end, as shown in \figref{fig:expand_merge(a)}, we continue to employ a training-free weight aggregation approach, instead of distillation, but we leverage the prediction of optimization directions for aggregating the cluster center model parameters to avoid a decline in accuracy.
This approach ensures that there are no extra costs on the client side, and model aggregation on the server side is more efficient.
In particular, we present Algorithm \ref{alg:merging} to illustrate the key steps:
In contrast to direct average aggregation, we distinguish between the main model, labeled as $v_{m}$, and the auxiliary model, denoted as $v_{aux}$. 
Empirically, we designate the model associated with a higher number of clients within the cluster as the main model $v_{m}$ (Line 1).
Then we extract valuable knowledge from $v_{aux}$ and incorporate them into $v_{m}$. 
Here, we treat the parameters of $v_{aux}$ as an optimization target for $v_{m}$ and calculate an optimization direction $\vec{v_{m}}$ (Line 2).
And we refine our optimization direction by incorporating the posterior distribution. 
\rev{Specifically, to evaluate the effectiveness of using $v_{aux}$ as an optimization target, we first perform local training on the main model $v_{m}$ using the local dataset $\Phi_{i}$, expressed as $v_{m}\textquotesingle\leftarrow F(v_{m},\Phi_{i})$(Line 3).} 
Subsequently, we calculate the weight difference between the model before training ($v_{m}$) and after training ($v_{m}\textquotesingle$). 
\rev{This process results in a posterior distribution $\vec{v_{m}\textquotesingle}$ for the optimization direction (Line 4).}
\rev{We leverage the weight-granularity attention matrix $\alpha$ in the refinement process(Line 5).} 
\rev{With this matrix, we can carry out weight-granularity aggregation on both $v_{m}$ and $v_{aux}$, resulting in the merged outcome $\Hat{v}$ (Line 6).}

\subsubsection{Mobile Cluster Expansion}
As shown in \figref{fig:expand_merge(b)}, during the expansion process, cluster expansion extends the client with the wrong cluster into a new cluster. 
If the client feedback is within the same cluster, it implies the current cluster would not fit all the mobile clients.
In this case, \sysname would split the cluster into two.
The server ranks the collected client feedback in ascending order.
If the client's feedback constitutes the last 20\%, it will be assigned to a new cluster $\hat{c}$. 
The client $\gamma$ assigned to the new cluster is removed from the original cluster $c$, while others remain in the original cluster. 

To rapidly learn the model parameters for the new cluster, we consider the new cluster as the original one with data drifts. 
Accordingly, we employ transfer learning that specifically targets domain adaptation to fine-tune the new cluster’s model parameters upon those of the original cluster (line 3).
However, the newly expanded cluster model obtained via transfer learning often suffers from overfitting issues~\cite{liu2022improved}. 
This is because a scarcity of data samples from the new data drifts can lead to overfitting of the cluster center model to a small amount of new data, thereby reducing model generalization. 
To tackle this issue, we propose that each mobile client in a newly expanded cluster conducts local training with partial fine-tuning, focusing on adjusting the final layer output rather than full training.
This partial fine-tuning restriction will only be lifted after the next cluster merging refinement, allowing the transition to normal full training mode.
Technically, we assign a boolean index to label each client's local training mode within the newly expanded cluster, indicating whether they should perform partial fine-tuning or full training.

%% file: body/broadcast.tex
\section{In-cluster model broadcast}
\label{sec:broadcast}
After client clustering, the server disseminates the latest models to clients engaged in training the same personalized model \ie within the same cluster, for staleness control.
Since client clustering effectively restrains the broadcast scope and mitigates the data heterogeneity, the core of in-cluster model broadcast is how to determine the broadcast frequency to further reduce the communication cost without compromising training accuracy.
First, we examine the benefits of model broadcast ( \secref{sec:broadcast_limitations}) for controlling staleness, aiming to prevent accuracy deterioration and slower convergence. Subsequently, we introduce our online broadcast frequency prediction scheme in \secref{sec:comm_topk}.

\subsection{Advantages of In-cluster Model Broadcast}
\label{sec:broadcast_limitations}
\tabref{tb_vs_transmit} compares the model dissemination strategy and the associated communication cost of different FL paradigms.
Existing FL paradigms, either synchronous or asynchronous, exhibit \textit{symmetric} client-server communication.
This is not aligned with the \textit{asymmetric bandwidth} in wireless networks, and thus \textit{under-utilization} of the \textit{downstream} bandwidth.
In contrast, \sysname takes advantage of the abundant downstream bandwidth to distribute the latest models to clients when necessary, preventing accuracy decline or convergence slowdown when the staleness problem is significant.
We empirically (\secref{sec:exp_comm}) show that the asymmetric server-client communication pattern in \sysname not only results in decreased overall communication cost, but also avoids severe communication peaks in existing FL paradigms \cite{mcmahan2017communication, lai2021oort, ouyang2021clusterfl, xie2019asynchronous}.

As a separate note, the staleness problem, as exemplified in \cite{sun2022fedsea}, always leads to reduced accuracy when aggregating out-of-date weights from stragglers.
This is a common issue because, in practical applications, the updates from mobile local models inevitably reach the server in different epochs.
Similar to~ \cite{koloskova2022sharper}, we represent the convergence rate for \sysname's asynchronous PFL  by $\mathcal{O}(\sqrt{Q_{max}Q_{avg}})$.
Where, $Q_{max}$ denotes the maximum staleness degree of models uploaded across the entire asynchronous PFL process, and $Q_{avg}$ signifies the average of that.
Aggregating a few outdated models can significantly increase $Q_{max}$, posing a bottleneck for convergence rate optimization.
Broadcast can effectively reduce $Q_{max}$ by distributing the cluster center model to clients, thereby improving the convergence rate.

\begin{table}[t]
\scriptsize
\caption{Comparison of model dissemination strategies and the communication cost of different FL paradigms.}
\vspace{-2mm}
\begin{tabular}{|c|c|c|c|}
\hline
\textbf{Approachs}         & \textbf{When ?}                     & \textbf{To whom ?}              & \textbf{Comm optimization}  \\ \hline
\textbf{FedAvg(Syn FL)}    & Wait for all devices                & Broadcast to all                & $\times$(high cost \& peak) \\ \hline
\textbf{Oort(Syn FL)}      & Wait for all devices                & Broadcast to all                & $\checkmark$(peak)          \\ \hline
\textbf{FedAsyn(Asyn FL)}  & Every local updates                 & Unicast to a device             & $\times$(high cost)         \\ \hline
\textbf{ClusterFL(Syn FL)} & Wait for all devices                & broadcast to a cluster          & $\times$(high cost \& peak) \\ \hline
\textbf{EchoPFL(Asyn PFL)}   & On-demand after every local updates & Adaptive broadcast to a cluster & $\checkmark$                \\ \hline
\end{tabular}
\label{tb_vs_transmit}
\vspace{-4mm}
\end{table}

\subsection{Online In-cluster broadcast Frequency Prediction}
\label{sec:comm_topk}
Within each cluster, \sysname finetunes the broadcast frequency to further balance the communication cost and model accuracy.
Rather than broadcast at fixed intervals or rounds, \sysname decides when to broadcast dynamically.

\subsubsection{RNN-based broadcast Frequency Predictor}
\label{sec:multicast_rnn}
Our strategy, termed \textit{model broadcast}, distributes the latest model to clients within the same cluster, proactively controlling model staleness.
The broadcast scope is automatically managed by client clustering algorithms.
The \textit{broadcast frequency} is \textit{dynamically set}.
Specifically, broadcast decisions are made after each model aggregation, by comparing the \textit{accumulated} model changes since the last broadcast and predicted model change following the next model aggregation.
\rev{Broadcast is invoked when the predicted model change exceeds the accumulated model changes, \ie $L_1(\hat{v}^{(t+1)}, v^{(t)})$ > $L_1(v^{(t)}, v_{m}^{(t)})$, where $\hat{v}^{(t+1)}$, $v^{(t)}$, and $v_{m}^{(t)}$ are the predicted aggregated model at time $t+1$, the aggregated model at time $t$, and the last broadcast model till time $t$, respectively.}
The rationale is that model changes between successive aggregations diminish in convergent training \cite{mcmahan2017communication, li2019convergence}. 
A substantial model change indicates intolerable model staleness.
The aggregation of such models into the cluster center models has become a bottleneck in minimizing the maximum staleness $Q_{max}$, which is known to slow down the convergence rate. 
Therefore, model broadcasting becomes necessary.
For simplicity, we measure the model changes in L1 distance and adopt a naive recurrent neural network to predict \rev{$\hat{v}^{(t+1)}$} based on historical models after each aggregation.

This method allows us to capture the most recent and substantial updates.
$K$ is proportional to the current number of clients within the cluster.
To save storage, we keep the change degree (\eg L1-distance) of these $K$ models rather than their model parameters.

We represent the set of these $K$ records at each round $t$ as $\mathbb{K}(t)$.
We input these Top-K alterations sequentially into an RNN model to decide whether to broadcast for a cluster.
The input length is the number of clients in the cluster. 
\rev{The RNN model has two hidden layers, each with 128 units. We use 1,200 historical states to pre-train
the RNN model and the last $K$ states for online fine-tuning.}
The training loss in RNN is:
\begin{equation}
\begin{split}
& loss = CrossEntropy(\mathcal{P}(\mathbb{K}(t-1)), \mathcal{G}(\mathbb{K}(t)))
\\
        & \mathcal{G}(\mathbb{K}(t)) =  
\begin{cases}
    1 (\text{broadcast}), & \text{if } h(v_{c}^{t-1},v_{c}^{t},v_{broadcast}^{t-1}) \geq 0 \\
    0 (\text{not broadcast}), & \text{if } h(v_{c}^{t-1},v_{c}^{t},v_{broadcast}^{t-1}) < 0
\end{cases}
\\
& h(v_{c}^{t-1},v_{c}^{t},v_{broadcast}^{t-1}) =L1(v_{c}^{t-1},v_{broadcast}^{t-1})-L1(v_{c}^{t-1},v_{c}^{t})
\end{split}
\end{equation}
where $\mathcal{P}(\mathbb{K}(t-1))$ is the prediction made by the RNN-based predictor on the historical records of $K$ alterations after communication round $(t-1)$.
$\mathcal{G}(\mathbb{K}(t))$ is the ground truth at $t-th$ round, which is obtained by computing the L1 distance between the weights of the previously broadcasted model $v_{broadcast}^{t-1}$ and the newly aggregated model $v_{c}^{t-1}$ before $t-th$ round (\ie the accumulated gap in model staleness), as well as the L1 distance between the newly aggregated model $v_{c}^{t-1}$ and the next iteration's aggregated model $v_{c}^{t}$ (\ie the eliminated model staleness).
This formulation dynamically balances broadcast frequency and model accuracy. 
Consequently, \sysname broadcasts more frequently given notable model changes; and less frequently otherwise.

\subsubsection{Dynamic Predictor Maintenance}
\label{sec:multicast_predictor}
Since \sysname dynamically expands and merges clusters, it is crucial to continuously refine the Top-K records, RNN models, and broadcast strategies for each evolving cluster.
\sysname maintains the necessary states for the predictors of each cluster as follows.
\begin{itemize}
\item \textit{Predictor Maintenance in Cluster Expansion.} 
\begin{itemize}
    \item The expanded cluster \textit{resets Top-K historical records} as the newly expanded client and pads zeros since existing historical records are inapplicable to the new cluster.
    \item The expanded cluster \textit{inherits} the RNN model weights as the initial weights.
    \item \textit{broadcast} is deactivated since the cluster center weight is already up-to-date before expansion.
\end{itemize}
\item \textit{Predictor Maintenance in Cluster Merging.} 
\begin{itemize}
    \item The merged cluster \textit{refreshes Top-K records} by sampling distinct records from each cluster before merging. 
    The sampling ratio is set proportional to the variance of the cached $K$ L1 distances in each cluster. 
    It prioritizes the selection of Top-K records with larger weight changes.
    \item We adopt knowledge distillation to merge the RNN weights of multiple clusters as \secref{subsec:cluster_merging}. 
    \item The merged cluster model is \textit{immediately broadcast} to its clients because cluster merging induces drastic weight changes and thus model staleness.
\end{itemize}
\end{itemize}
In summary, the online predictor, coupled with the flexibility to adjust predictors and the value of $K$, empowers \sysname to adapt to heterogeneous and dynamic mobile scenarios.

%% file: body/Implementation.tex
\section{CI-based Version Control Implementation}
\label{sec:implementation}


\begin{figure}[t]
  \centering
  \includegraphics[width=0.75\textwidth]{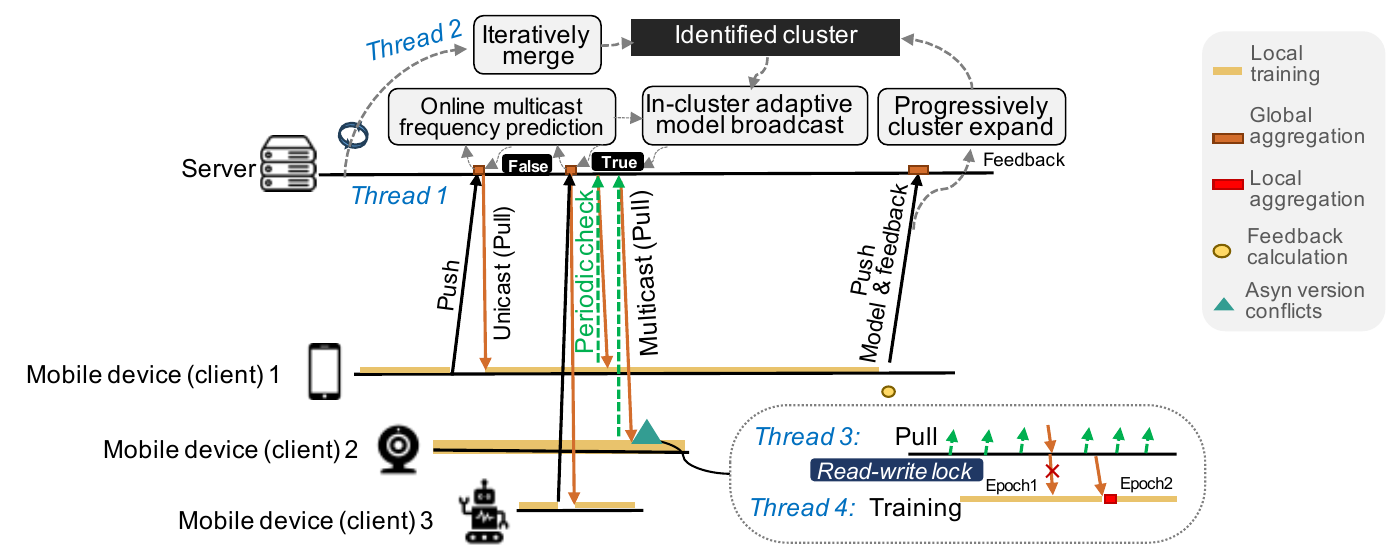}
  \vspace{-10pt}
  \caption{Continuous integration-based client-server asynchronous coordination mechanism.}
  \vspace{-10pt}
  \label{fig:workflow_feedback}
\end{figure}
\rev{In the APFL system, it is common for multiple mobile clients to
update the global model simultaneously, or for new global
model updates to be sent during local training, which could
lead to version conflicts. We draw inspiration from the continuous integration (CI) mechanism in Git, to implement
FedOM as an easy-to-use client-server coordination scheme
for integration with other FL frameworks. 
The CI system mechanism offers specific advantages for APFL's ubiquitous applications: i) Conflict resolution: The CI mechanism
adeptly handles version conflicts by controlling the aggregation of multiple model updates. ii) Immediate server feedback:
Clients receive prompt feedback about their uploads, mitigating and addressing errors stemming from heterogeneities in later training stages and enhancing convergence. iii) Fast mobile application release: Mobile clients can promptly deploy services with newly updated models, minimizing the interval between global
model training and its availability to mobile users. iv) Efficient branch collaboration: CI facilitates efficient collaboration in different branches, which, in the APFL context, are clusters.
We present the following operations, as shown in \figref{fig:workflow_feedback}:}

\begin{itemize}
    \item \textit{\textbf{\sysname Pull}: Fetch models as desired.}
    Clients periodically query the server for significant model changes and fetch the latest model from the server to the mobile device for synchronization.
    \item \textit{\textbf{\sysname Push}: Upload local updates on-demand.}
    Clients upload their local models to the server, and the server aggregates the received local models accordingly.
    \item \textit{\textbf{\sysname Branch}: Identify personalized clusters.} 
    \rev{Clustering is essential in personalized FL as it creates customized and targeted model updates for groups of clients with similar data characteristics. We use clustering to identify clusters of clients with similarity, and we implement these clusters' model updates as branches. We utilize multi-thread and read-write locks to resolve conflicts among personalized branches.}
\end{itemize}

%% file: body/experiment.tex
\section{Experiment}
\label{sec:experiment}

\begin{figure*}[t]
  \centering
  \subfloat[CIFAR-10]{
    \includegraphics[height=0.17\textwidth]{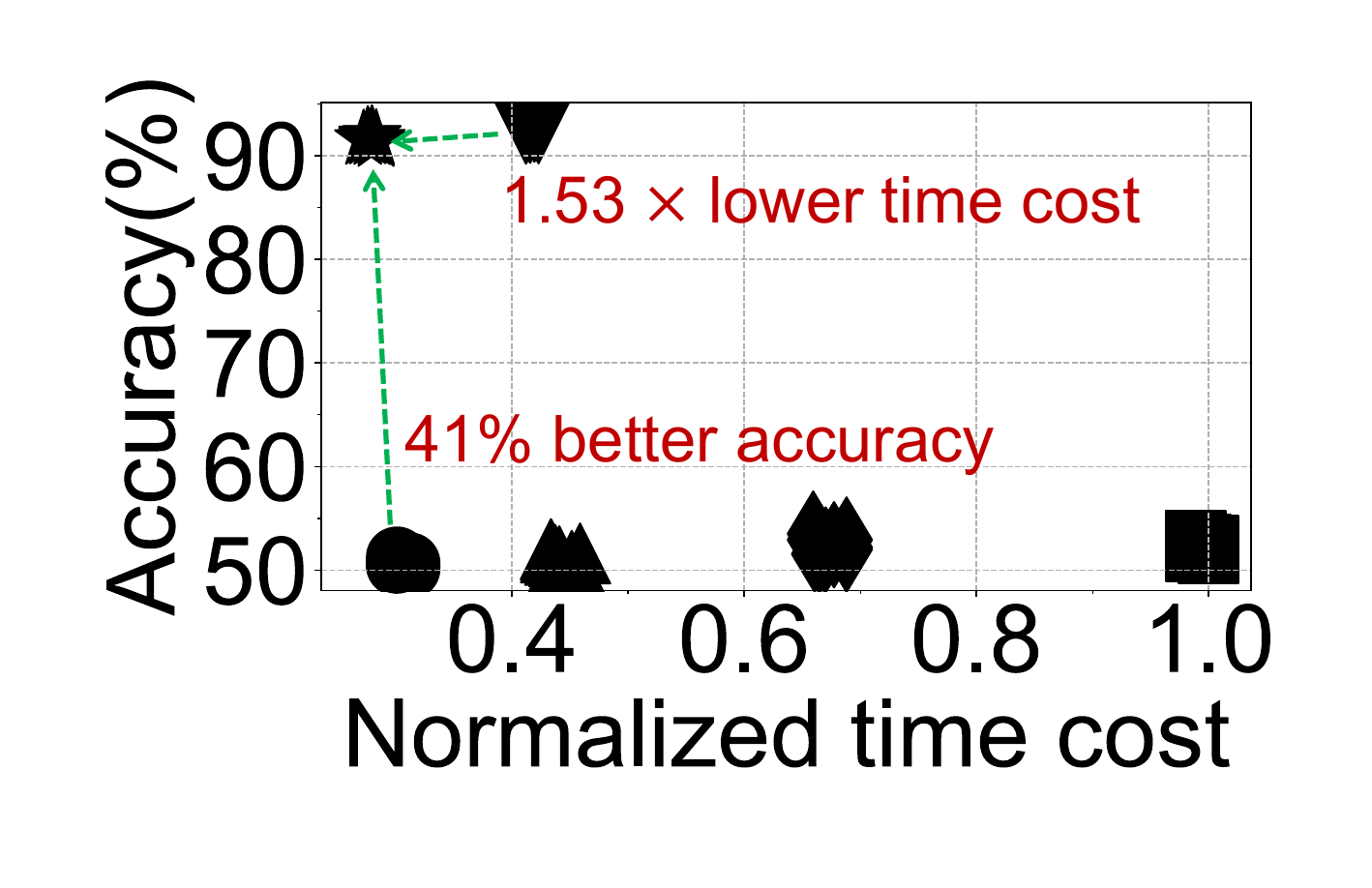}
    \vspace{-10pt}
    \label{fig:cifar_overall}
  }
  \hspace{2mm}
  \subfloat[HAR]{
    \includegraphics[height=0.17\textwidth]{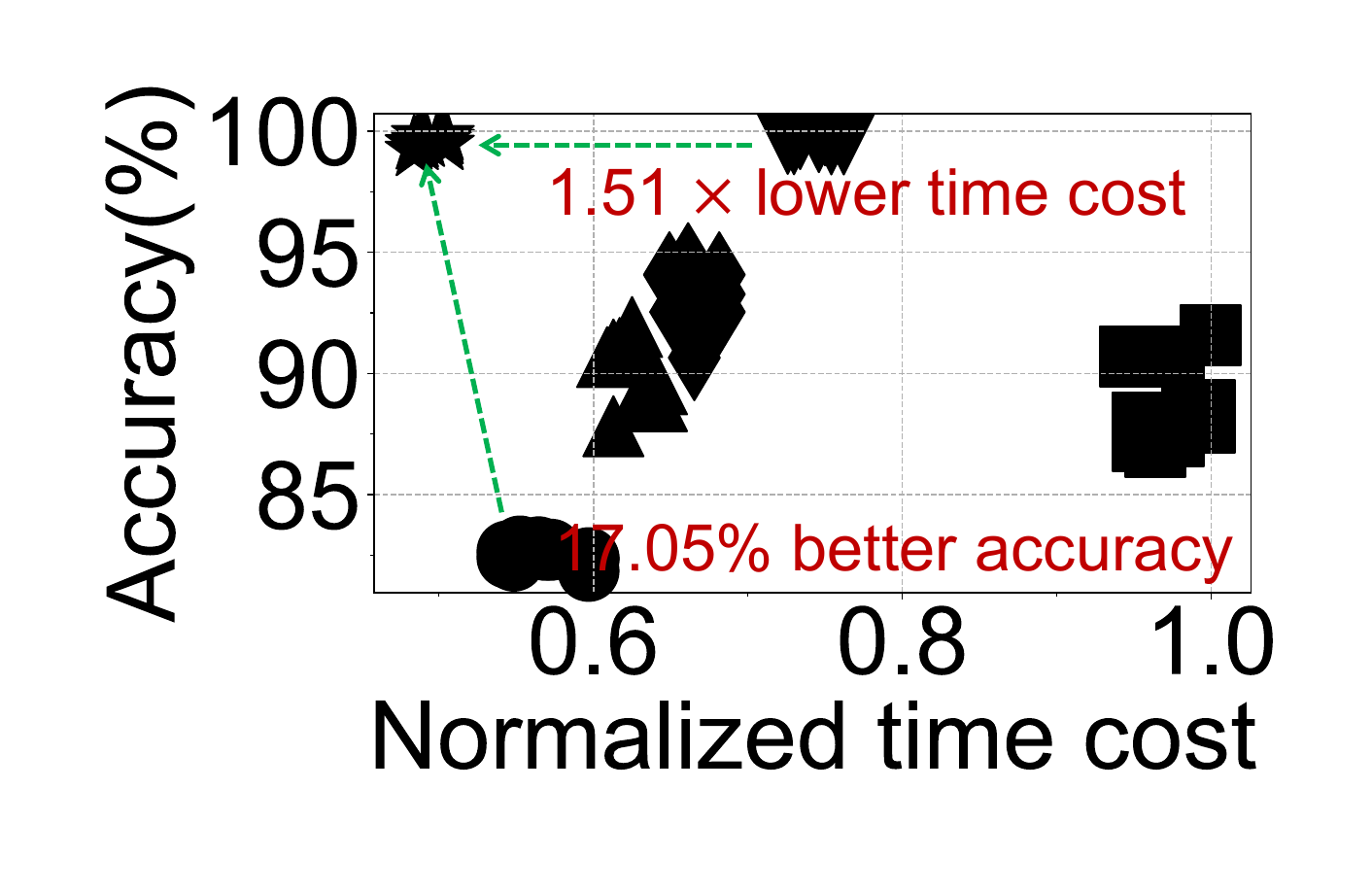}
    \vspace{-10pt}
    \label{fig:har_overall}
  }
  \hspace{2mm}
  \subfloat[Ubisound]{
    \includegraphics[height=0.17\textwidth]{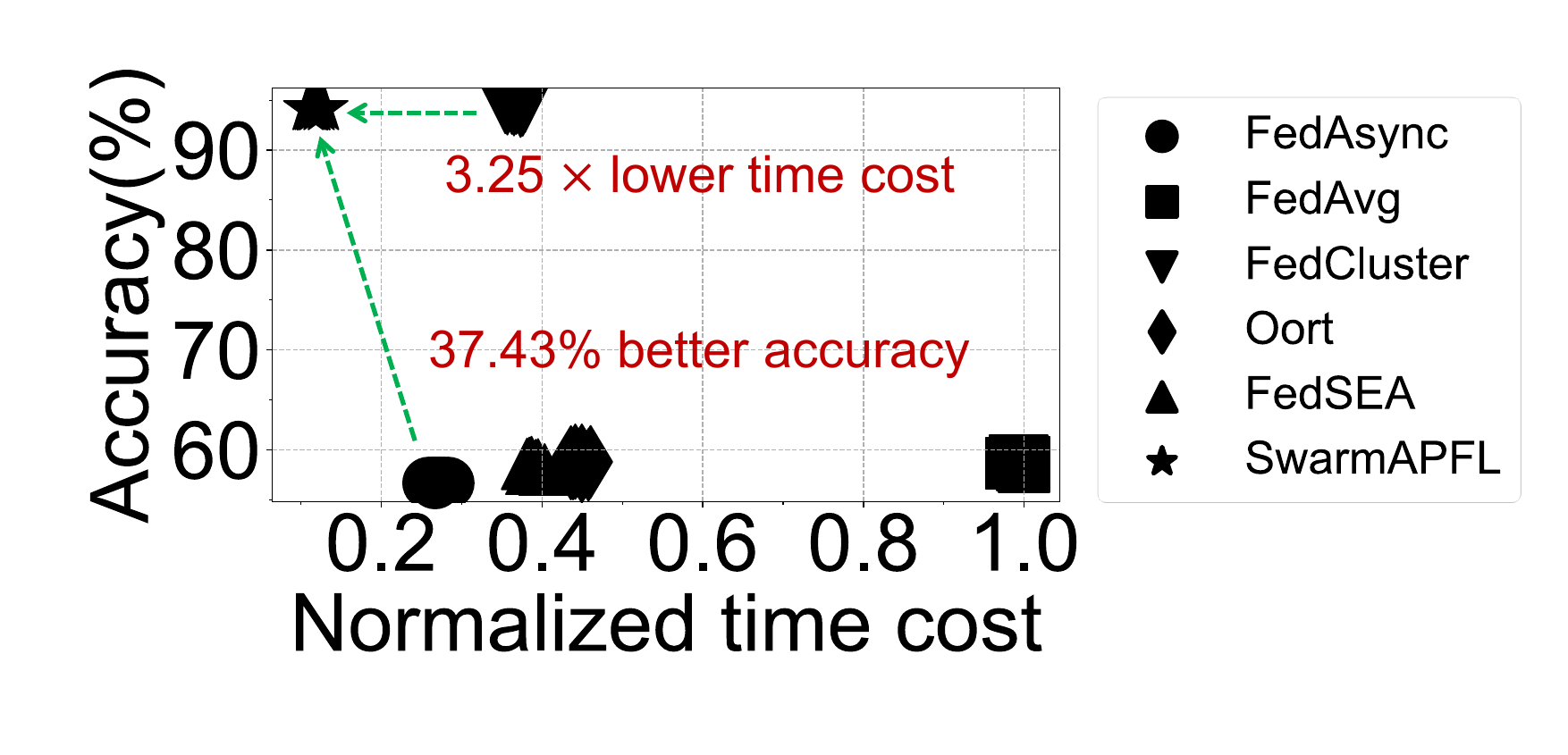}
    \vspace{-10pt}
    \label{fig:sound_overall}
  }
  \vspace{-10pt}
  \caption{Comparison of accuracy \textit{vs.} training time between \sysname and other baselines on diverse tasks.}
  \vspace{-10pt}
  \label{fig:baseline}
\end{figure*}

\subsection{Experiment Setup}
\label{sec:experiment:setup}

\fakeparagraph{Implementation}
We implement \sysname using Python 3.7 and PyTorch 1.10 for the server and mobile clients, respectively. 
The server is equipped with two RTX 3080 GPUs and 128GB RAM.
We use 20 mobile and embedded devices of five types: Jetson Nano ($D_1$), Jetson NX Xavier ($D_2$), Jetson Nano Orin ($D_3$), Jetson AGX Xavier ($D_4$), and Raspberry Pi 4 ($D_5$). 
They represent diverse computing capabilities and form a distributed FL system. 
\rev{
For the implementation of \sysname, we set the hyperparameter $C$ to 2 in \secref{sec:cluster_init}. 
For the RNN implementation in \secref{sec:multicast_rnn}, We use two hidden layers to construct the RNN model, with each layer consisting of 128 units. 
We pre-train the RNN model using $1,200$ historical states, and for online fine-tuning, we utilize the last $K$ states.
The $K$ value is set to 10.}

\fakeparagraph{Tasks, Datasets, and Models}
We experiment with four real-world mobile applications.
And the data assigned to each client is Non-IID and unbalanced.
\begin{itemize}
    \item \textbf{Image Recognition} ($T_1$) is ubiquitous in smart cameras/robots. 
    We employed the CIFAR-10 dataset \cite{krizhevsky2009learning}. \rev{For Non-IID setting, each device contains 2-class data, and the data within each class can be unbalanced.}
    The model has two convolutional (\texttt{conv}) layers followed by a fully connected (\texttt{fc}) layer.
    \item \textbf{Human Activity Recognition, HAR ($T_2$)} on mobiles has gained significant attention~\cite{zhang2023passive}. 
    We adopt the HAR-UCI dataset~\cite{anguita2013public}, which comprises sensor data from 30 users.
    The model contains two \texttt{fc} layers.
    \item \textbf{Sound Detection ($T_3$)} for hard-of-hearing people using wearables is crucial. 
    We use the Ubisound~\cite{sicong2017ubiear} dataset, comprising nine sound classes.
    \rev{For Non-IID setting, each device contains 3-class data, and the data within each class can be unbalanced.}
    The model contains two \texttt{conv} layers followed by two \texttt{fc} layers.
    \item \textbf{Automatic Image File Cleaning} ($T_4$) helps users mange image files.
    We collected a dataset of $15000$ images from phones, robots, and dashboard cameras. \rev{For Non-IID setting, devices hold unbalanced "Delete/Retain" data.}
    The model has two \texttt{conv} layers and two \texttt{fc} layers. 
\end{itemize}
\rev{Assessing the performance of the FL system can be challenging when dealing with hundreds of physical devices.
Therefore, we divide our experiments into simulation experiments and real-world experiments.
For simulation experiments($T_{1}$, $T_{2}$ and $T_{3}$), we gather data on local training times and communication overhead to simulate the system using software. In detail, we simulate 20\% $D_{1}$, 20\% $D_{2}$, 20\% $D_{3}$ and 40\% $D_{5}$. 
For real-world experiments($T_{4}$), we conducted them with 3$D_{1}$, 5$D_{2}$, 4$D_{3}$,2$D_{4}$ and 6$D_{5}$.}

\fakeparagraph{Baselines}
We adopt six mainstream FL algorithms with mobile devices as performance comparison baselines. 
They are configured as follows:
\begin{itemize}
    \item \textbf{Synchronous FL}: the server waits for all mobile clients for each round.
    It sets the accuracy baseline because it has the most comprehensive knowledge from all clients.
    It also sets a tough communication cost line due to its infrequent communication frequency.
    \begin{itemize}
        \item \textbf{FedAvg} ~\cite{mcmahan2017communication}: 
        The server calculates the average of all mobile clients' weights.
        The updated global model is then broadcast back to all clients.
        \item \textbf{Oort}~\cite{lai2021oort} adopts mobile client selection to reduce the waiting time due to system heterogeneity.
    \end{itemize}
    \item \textbf{Asynchronous FL: FedAsyn}~\cite{xie2019asynchronous} promptly aggregates the model and distributes updates to mobile clients in an asynchronous manner.
    \item \textbf{Semi-asynchronous FL: FedSEA}~\cite{sun2022fedsea} balances model accuracy and training latency through scheduling synchronization points. 
    It also optimizes the error caused by weight discarding of slow devices.
    \item \textbf{Synchronous PFL: ClusterFL}~\cite{ouyang2021clusterfl} 
    trains multiple personalized models by clustering clients based on the similarity of their model outputs. 
    And it identifies personalized clusters based on their similarities.
    \item \textbf{Asynchronous PFL: \sysname} integrates PFL into the asynchronous framework.
     \item \textbf{Standalone}: involves individual training on each client, without federated training with other clients.
\end{itemize}

\subsection{Performance Comparison}
\label{ex_compare}

\subsubsection{Model Accuracy vs. Training Latency}
We test three typical mobile tasks: image recognition ($T_1$), activity recognition ($T_2$), and sound detection ($T_3$).
\rev{In this experiment, we use the simulation environment including 20\% $D_{1}$, 20\% $D_{2}$, 20\% $D_{3}$ and 40\% $D_{5}$.}
\figref{fig:baseline} shows the results. 
First, \sysname exhibits the best balance between model accuracy and training time compared to the baselines. 
Second, \sysname outperforms the four non-personalized baselines in accuracy.
It achieves comparable accuracy to ClusterFL, the state-of-the-art PFL method, across all three tasks. 
For instance, on Ubisound, \sysname demonstrates accuracy improvements of 37.4\%, 35.4\%, 35.8\%, and 35.6\% compared to FedAsyn, FedAvg, Oort, and FedSEA, respectively. 
Third, \sysname consistently yields the lowest training time across all three tasks. 
For CIFAR-10, its training was up to 3.7$\times$ faster than the baselines. 
This efficiency enhancement during convergence is attributed to personalized FL's capacity to forego convergence directions irrelevant to diverse personalized models.

\textbf{Summary.} 
\sysname achieved the best overall trade-off between model accuracy and training time.
This makes \sysname a promising solution for federated learning in the presence of data and system heterogeneity in ubiquitous mobile applications.

\subsubsection{Communication Cost}
\label{sec:exp_comm}
We test the communication efficiency of \sysname despite its broadcast strategy.
\begin{figure}[!t]
  \centering  \includegraphics[width=0.48\textwidth]{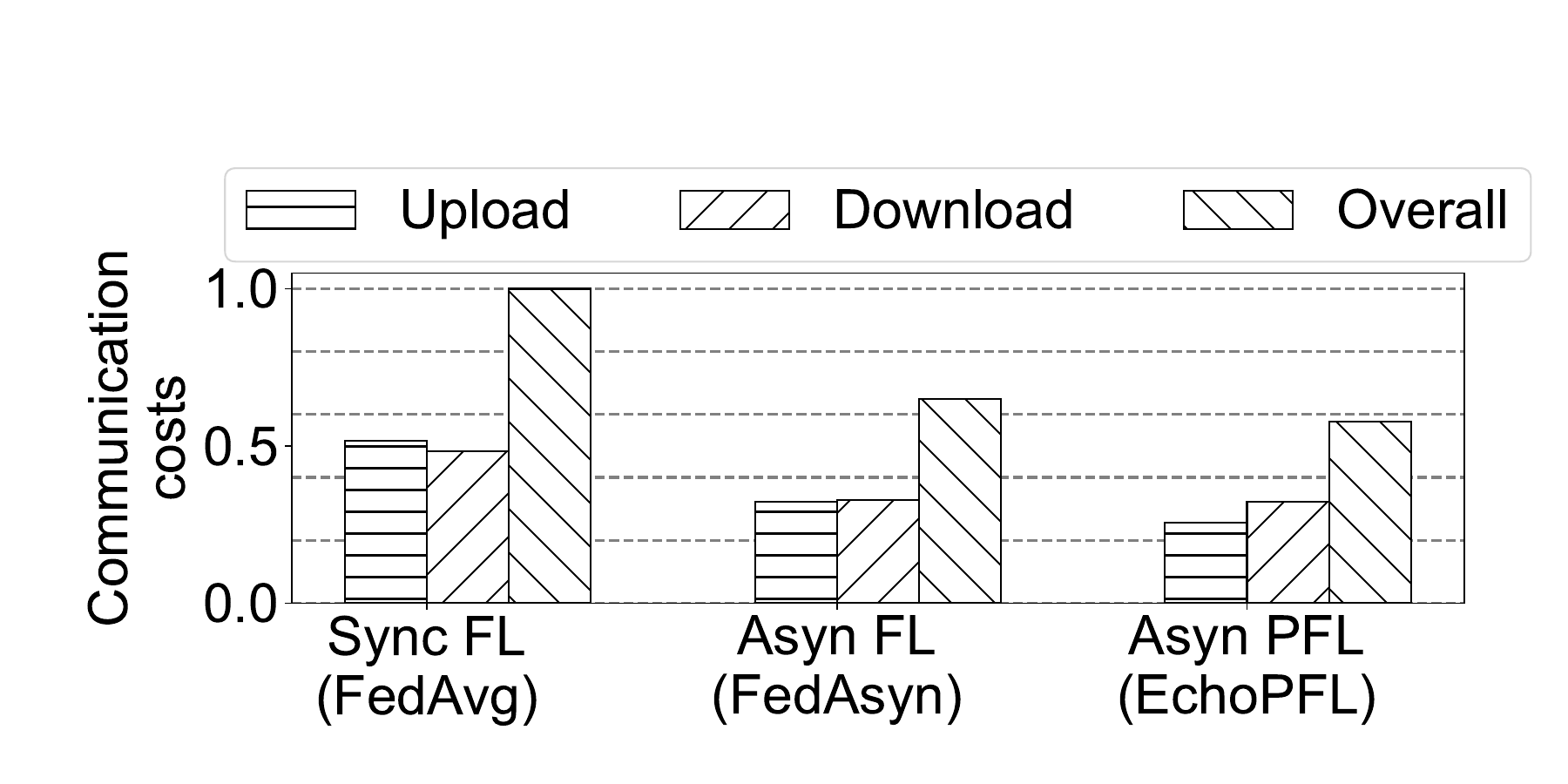}
  \vspace{-10pt}
  \caption{\sysname \textit{vs.} other baselines in terms of upload, download, and overall communication cost.}
  \vspace{-15pt}
  \label{fig:comm_cost}
\end{figure}
\begin{figure}[!t]
  \centering
  \subfloat[Download]{
    \includegraphics[width=0.42\textwidth]{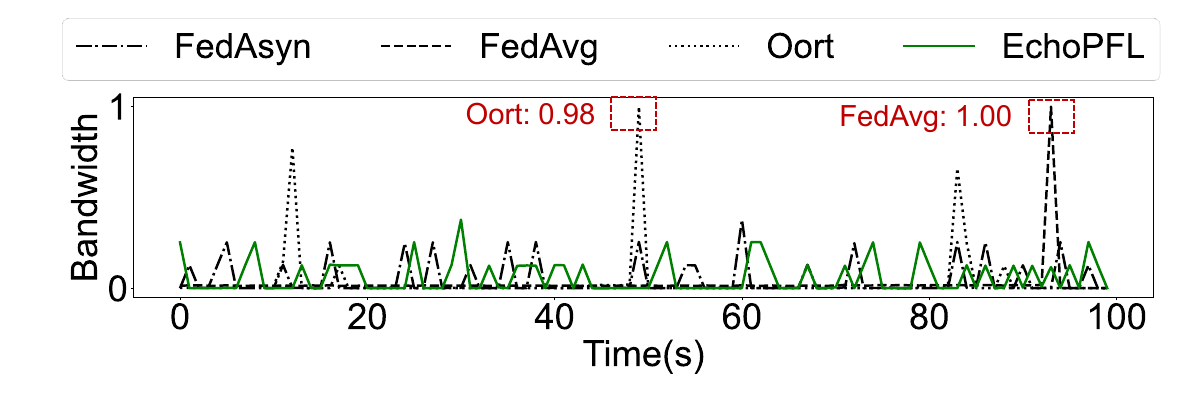}
    \label{fig:comm_curve(a)}
    \vspace{-10pt}
  }
  \hfill
  \subfloat[Upload]{
    \includegraphics[width=0.48\textwidth]{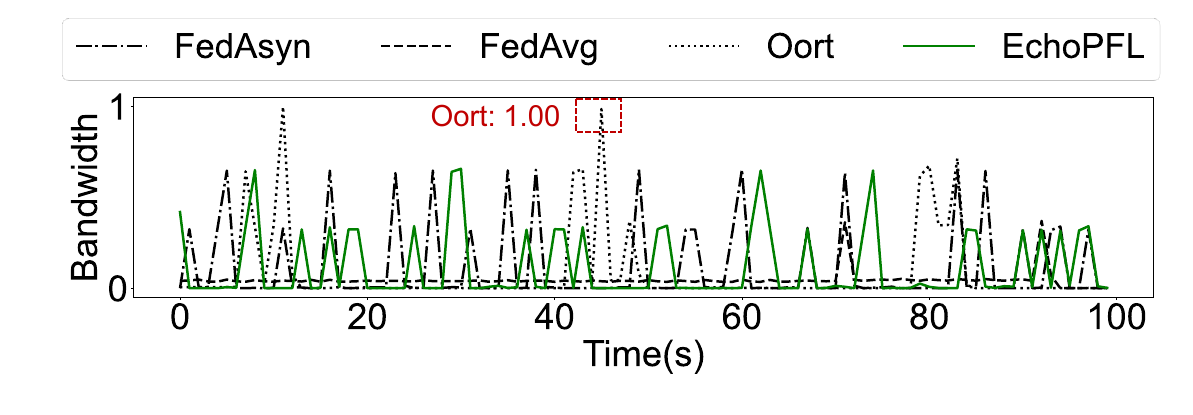}
    \label{fig:comm_curve(b)}
    \vspace{-10pt}
  }
  \vspace{-10pt}
  \caption{Communication curve in (a) download and (b) upload processes.}
  \vspace{-10pt}
  \label{fig:comm_curve}
\end{figure}

\begin{table}[t]
\scriptsize
\caption{Communication frequency and training time.}
\vspace{-10pt}
\begin{tabular}{c|c|c}
\hline
        & \begin{tabular}[c]{@{}c@{}}\textbf{Communication frequency} (per minute upload)\end{tabular} & \begin{tabular}[c]{@{}c@{}}\textbf{Training time}  (min)\end{tabular} \\ \hline
\textbf{Syn FL (FedAvg)}  & 29.22                                                                           & 398                                                              \\ \hline
\textbf{Asyn FL (FedAsyn)} & 64                                                                            & 102                                                              \\ \hline
\textbf{EchoPFL}    & 61.8                                                                            & 81                                                               \\ \hline
\end{tabular}
\label{tb_communication_2}
\end{table}

\begin{itemize}
    \item \textit{Overall Communication Cost.}
    We compare \sysname's communication cost with FedAvg, FedAsyn, and FedSEA in the image recognition ($T_{1}$) task. 
    \rev{The models are deployed on five clients: one $D_1$, two $D_2$, and two $D_4$.}
    We focus on the sum of upload and download communication costs.
    As depicted in \figref{fig:comm_cost}, \sysname achieves a $37\%$ reduction in overall communication costs compared to FedAvg, a $25\%$ reduction compared to FedSEA, and comparable communication costs to FedAsyn.
    \item \textit{Upload and Download Communication Cost.}
    To understand why \sysname reduces the overall communication cost, we assess the specific costs associated with upload and download communications.
    As shown in \figref{fig:comm_curve} and \tabref{tb_communication_2}, 
    \sysname reduces training time by $79.6\%$, significantly decreasing overall communication cost, even though it results in a $3.12 \times$ higher download frequency than FedAvg.
    Compared to FedAsyn, \sysname exhibits $1.42 \times$ more download frequency but achieves a $20.6\%$ lower convergence time. 
    This indicates that increasing the download frequency does not increase the training time; instead, it brings faster convergence.
    \item \textit{Communication Peak.}
    We compared communication peaks between \sysname and three baselines over a monitoring duration of 1 hour. 
    Communication peaks can often result in packet loss and communication disruptions.
    As illustrated in \figref{fig:comm_curve}, both FedAvg and Oort exhibit frequent communication peaks, mainly due to their short-term synchronous download strategies. 
    In contrast, \sysname maintains relatively stable upload and download communication costs. 
    Specifically, the upload communication peak in \sysname is $1.48\times{}$ lower than that of Oort and $2.08 \times{}$ lower than FedAvg. 
    This difference is attributed to \sysname's prevention of large-scale simultaneous model distribution, which is a common characteristic of synchronous methods like FedAvg and Oort (see \figref{fig:comm_curve}(a)).
\end{itemize}
In summary, \sysname establishes a bandwidth-friendly solution for mobile FL with several advantages:
\textit{i)} 
\sysname boasts the lowest overall communication cost compared to FedAvg, FedAsyn, and FedSEA, attributed to its in-cluster broadcast method, which accelerates convergence and reduces the number of communication rounds.
\textit{ii)} 
While \sysname experiences higher download costs, it significantly curtails the upload cost, aligning well with the characteristics of asymmetrical wireless networks.
\textit{iii)} 
\sysname effectively eliminates communication peaks observed in synchronous FL.


\begin{figure}[t]
  \centering

  \subfloat[Initial state]{
    \includegraphics[width=0.2\textwidth]{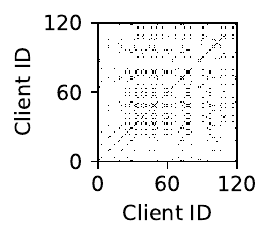}
    \vspace{-10pt}
    \label{fig:cluster_result(a)}
  }
  \hspace{5mm}
  \subfloat[\sysname]{
    \includegraphics[width=0.2\textwidth]{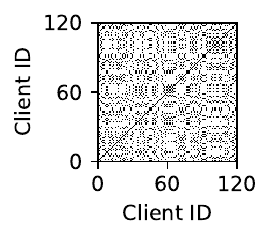}
    \vspace{-10pt}
    \label{fig:cluster_result(b)}
  }
  \hspace{5mm}
  \subfloat[ClusterFL]{
    \includegraphics[width=0.2\textwidth]{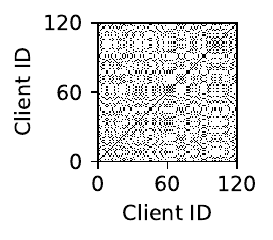}
    \vspace{-10pt}
    \label{fig:cluster_result(c)}
  }
  \vspace{-10pt}
  \caption{Comparing \sysname with ClusterFL~\cite{ouyang2021clusterfl} in clustering results among 120 clients, visualized by a boolean matrix of collaboration relationships.} 
  \vspace{-10pt}
  \label{fig:cluster_result}
\end{figure}

\begin{figure}[t]
  \centering
  \subfloat[Cluster similarity]{
    \includegraphics[width=0.25\textwidth]{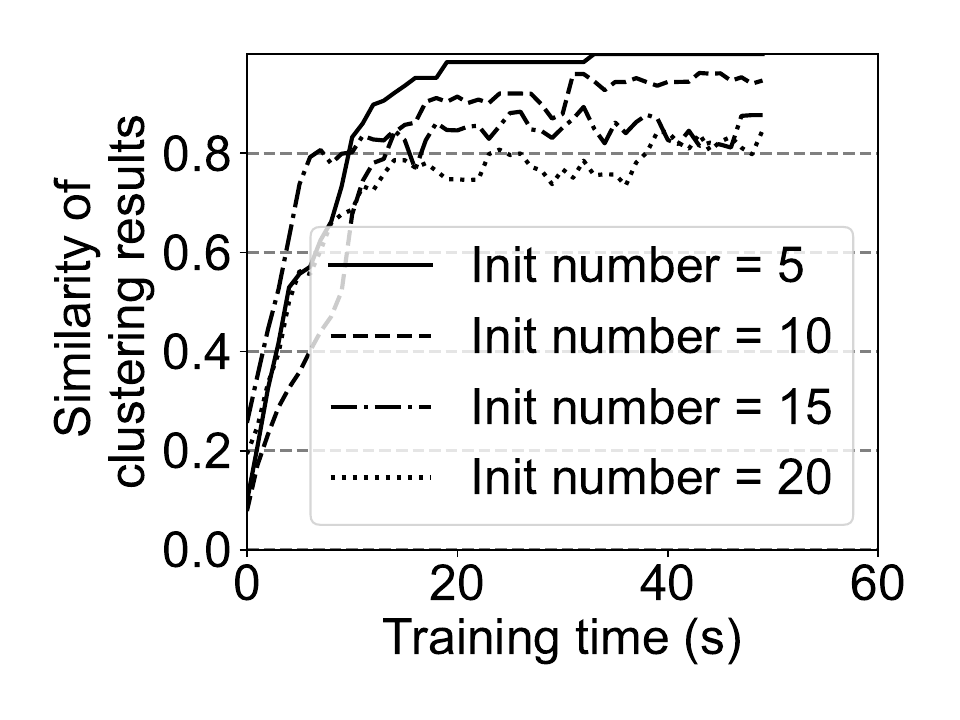}
    \vspace{-10pt}
    \label{fig:cluster_end_result(a)}
  }
  \hspace{8mm}
  \subfloat[Time to accuracy]{
    \includegraphics[width=0.25\textwidth]{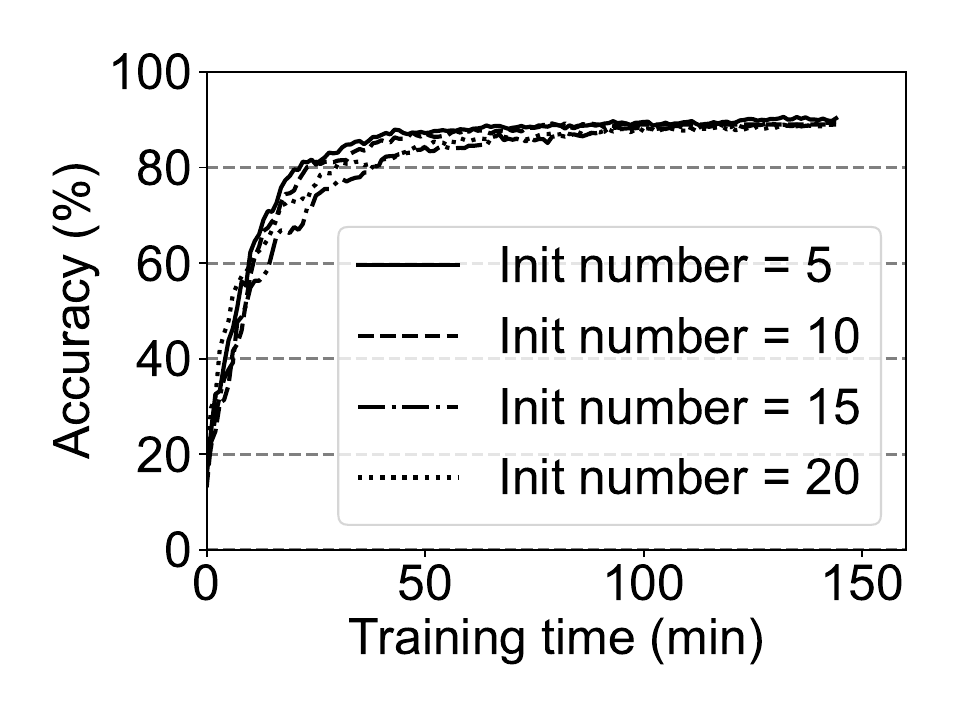}
    \vspace{-10pt}
    \label{fig:cluster_end_result(b)}
  }
  \vspace{-10pt}
  \caption{The impact of the initial number of clusters.}
  \vspace{-10pt}
  \label{fig:cluster_end_result}
\end{figure}

\begin{table*}[t]
\scriptsize
\caption{Experimental data and device heterogeneity settings in four real-world scenarios.}
\begin{tabular}{|c|cc|cc|cc|cc|cc|}
\hline
\multirow{2}{*}{\textbf{NO.}} & \multicolumn{2}{c|}{\textbf{Client 1}} & \multicolumn{2}{c|}{\textbf{Client2}} & \multicolumn{2}{c|}{\textbf{Client3}} & \multicolumn{2}{c|}{\textbf{Client4}} & \multicolumn{2}{c|}{\textbf{Client5}} \\ \cline{2-11} 
 & \multicolumn{1}{c|}{\textbf{Device}} & \textbf{Data} & \multicolumn{1}{c|}{\textbf{Device}} & \textbf{Data} & \multicolumn{1}{c|}{\textbf{Device}} & \textbf{Data} & \multicolumn{1}{c|}{\textbf{Device}} & \textbf{Data} & \multicolumn{1}{c|}{\textbf{Device}} & \textbf{Data} \\ \hline
\textbf{ A} & \multicolumn{1}{c|}{$D_1$} & 10\%class 1$\sim$ 10 & \multicolumn{1}{c|}{$D_1$} & 10\%class 1$\sim$ 10 & \multicolumn{1}{c|}{$D_1$} & 10\%class 1$\sim$ 10 & \multicolumn{1}{c|}{$D_1$} & 10\%class 1$\sim$ 10 & \multicolumn{1}{c|}{$D_1$} & 10\%class 1$\sim$ 10 \\ \hline
\textbf{ B} & \multicolumn{1}{c|}{$D_1$} & 25\%class1$\sim$ 4 & \multicolumn{1}{c|}{$D_1$} & 25\%class1$\sim$ 4 & \multicolumn{1}{c|}{$D_1$} & 50\%class1$\sim$ 2 & \multicolumn{1}{c|}{$D_1$} & 50\%class1$\sim$ 2 & \multicolumn{1}{c|}{$D_1$} & 25\%class1,75\%class 2 \\ \hline
\textbf{ C} & \multicolumn{1}{c|}{$D_1$} & 10\%class 1$\sim$ 10 & \multicolumn{1}{c|}{$D_1$} & 10\%class 1$\sim$ 10 & \multicolumn{1}{c|}{$D_2$} & 10\%class 1$\sim$ 10 & \multicolumn{1}{c|}{$D_2$} & 10\%class 1$\sim$ 10 & \multicolumn{1}{c|}{$D_4$} & 10\%class 1$\sim$ 10 \\ \hline
\textbf{ D} & \multicolumn{1}{c|}{$D_1$} & 25\%class1$\sim$ 4 & \multicolumn{1}{c|}{$D_1$} & 25\%class1$\sim$ 4 & \multicolumn{1}{c|}{$D_2$} & 50\%class1$\sim$ 2 & \multicolumn{1}{c|}{$D_2$} & 50\%class1$\sim$ 2 & \multicolumn{1}{c|}{$D_4$} & 25\%class1,75\%class 2 \\ \hline
\end{tabular}
\label{exp:scenario_tab}
\end{table*}

\begin{figure*}[t]
  \centering
  \subfloat[Scenario A: Homogeneous data and device]{
    \includegraphics[width=0.49\textwidth]{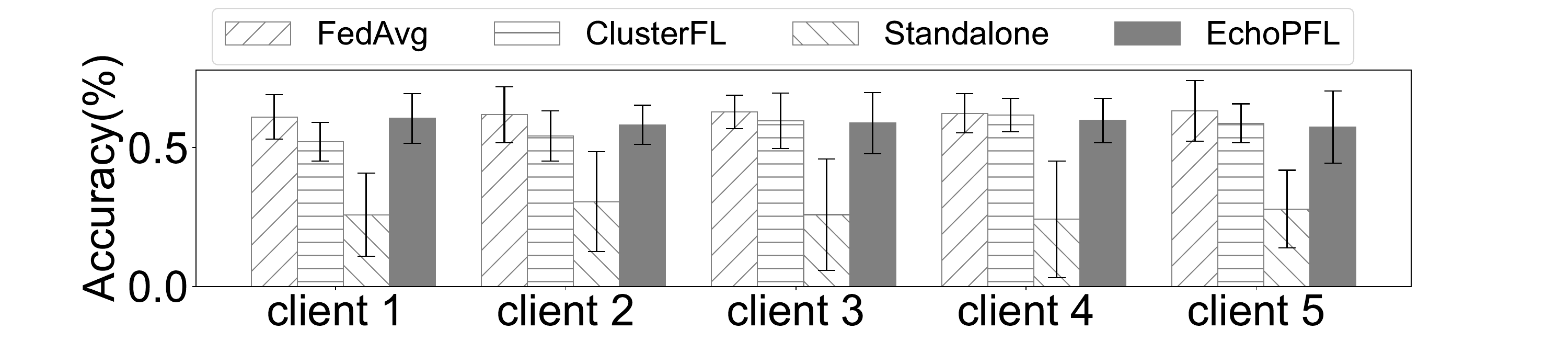}
    \vspace{-10pt}
    \label{fig:scenario_a}
    }
  \subfloat[Scenario B: Heterogeneous data]{
    \includegraphics[width=0.49\textwidth]{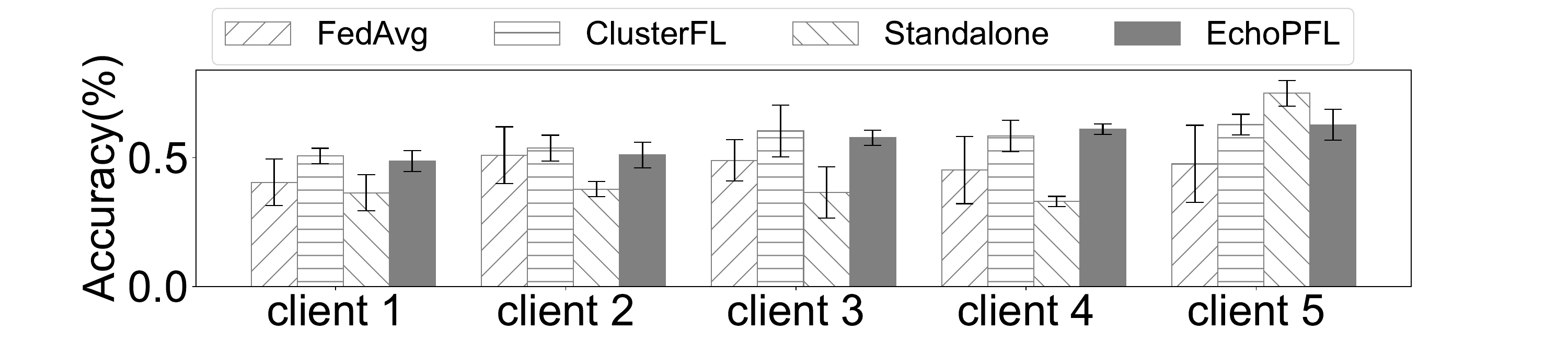}
    \vspace{-10pt}
    \label{fig:scenario_b}
  }
  \\
  \subfloat[Scenario C: Heterogeneous device]{
    \includegraphics[width=0.49\textwidth]{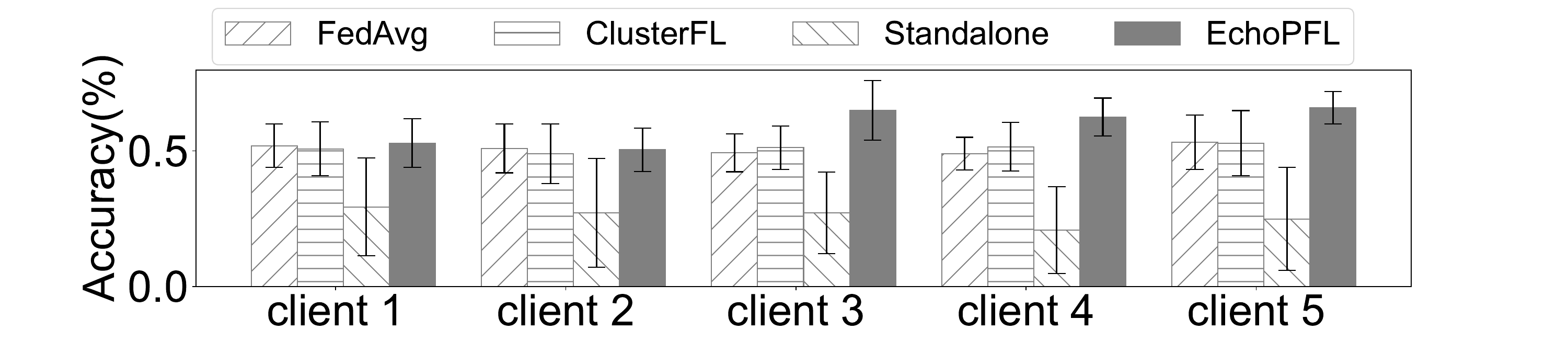}
    \vspace{-10pt}
    \label{fig:scenario_c}
  }
  \subfloat[Scenario D: Heterogeneous data and device]{
    \includegraphics[width=0.49\textwidth]{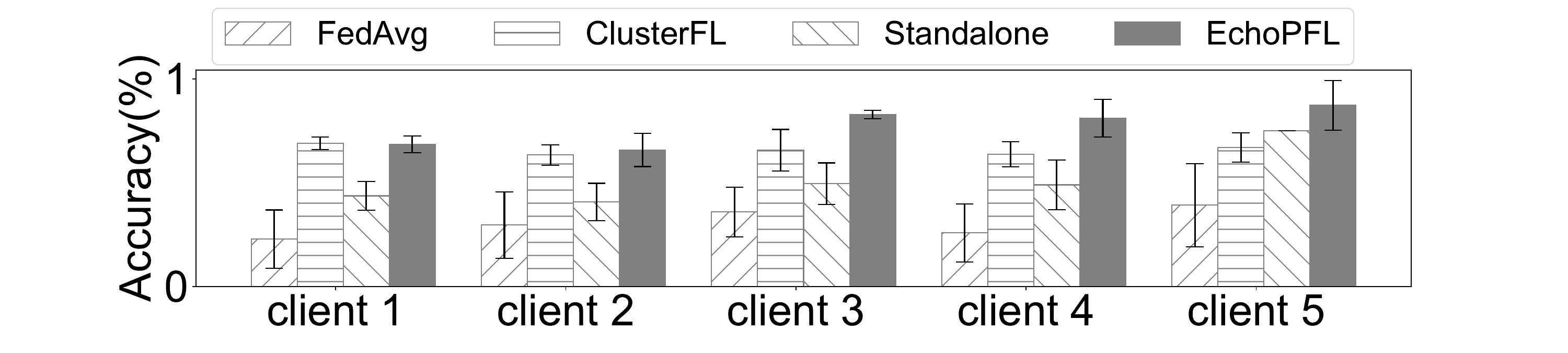
    }
    \vspace{-10pt}
    \label{fig:scenario_d}
  }
  \vspace{-10pt}
  \caption{Performance in four real-world scenarios.}
  \vspace{-10pt}
  \label{fig:scenario}
\end{figure*}

\begin{figure*}[t]
  \centering
  \subfloat[Sce A]{
    \includegraphics[height=0.15\textwidth]{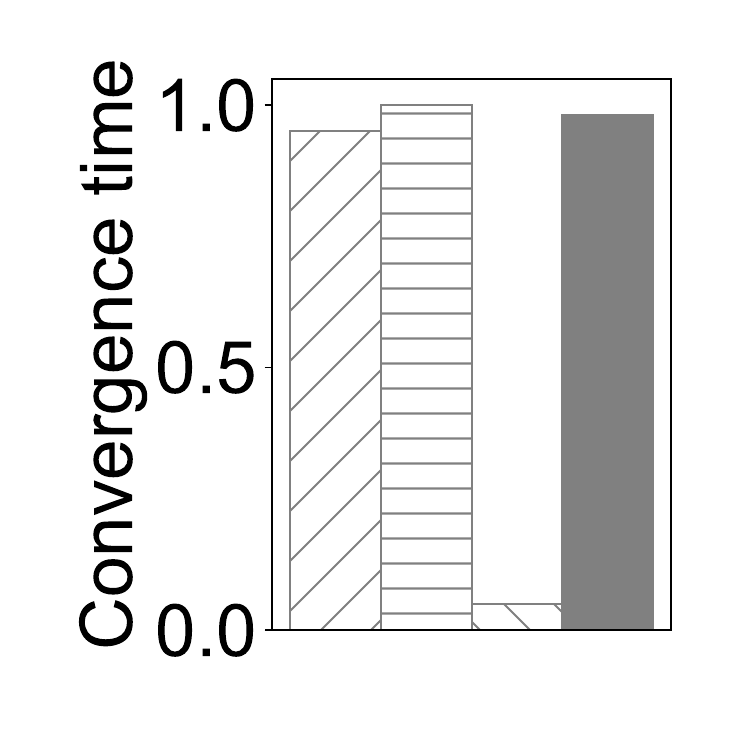}
    \vspace{-10pt}
    \label{fig:scenario_a_time}
  }
  \hspace{6mm}
  \subfloat[Sce B]{
    \includegraphics[height=0.15\textwidth]{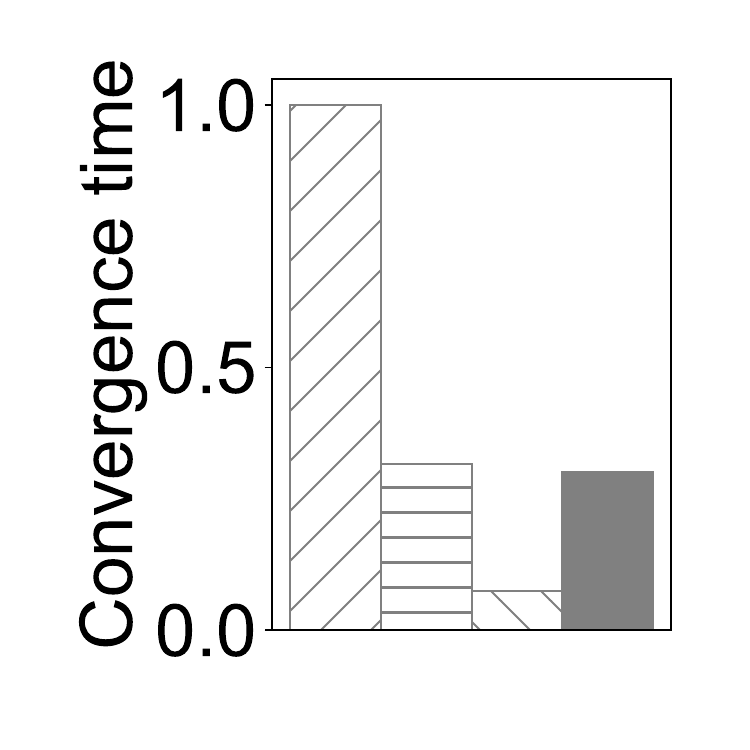}
    \vspace{-10pt}
    \label{fig:scenario_b_time}
  }
  \hspace{6mm}
  \subfloat[Sce C]{
    \includegraphics[height=0.15\textwidth]{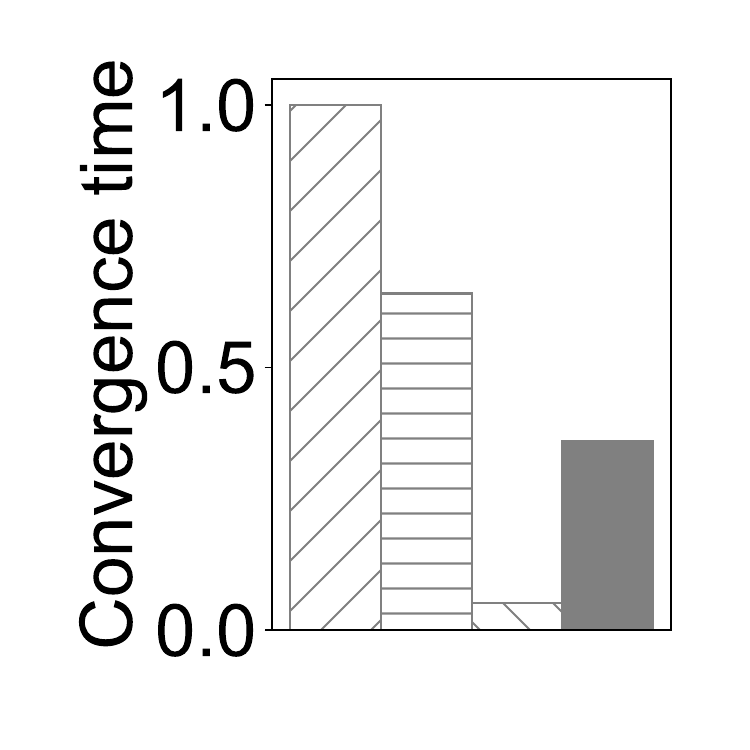}
    \vspace{-10pt}
    \label{fig:scenario_c_time}
  }
  \hspace{6mm}
  \subfloat[Sce D]{
    \includegraphics[height=0.15\textwidth]{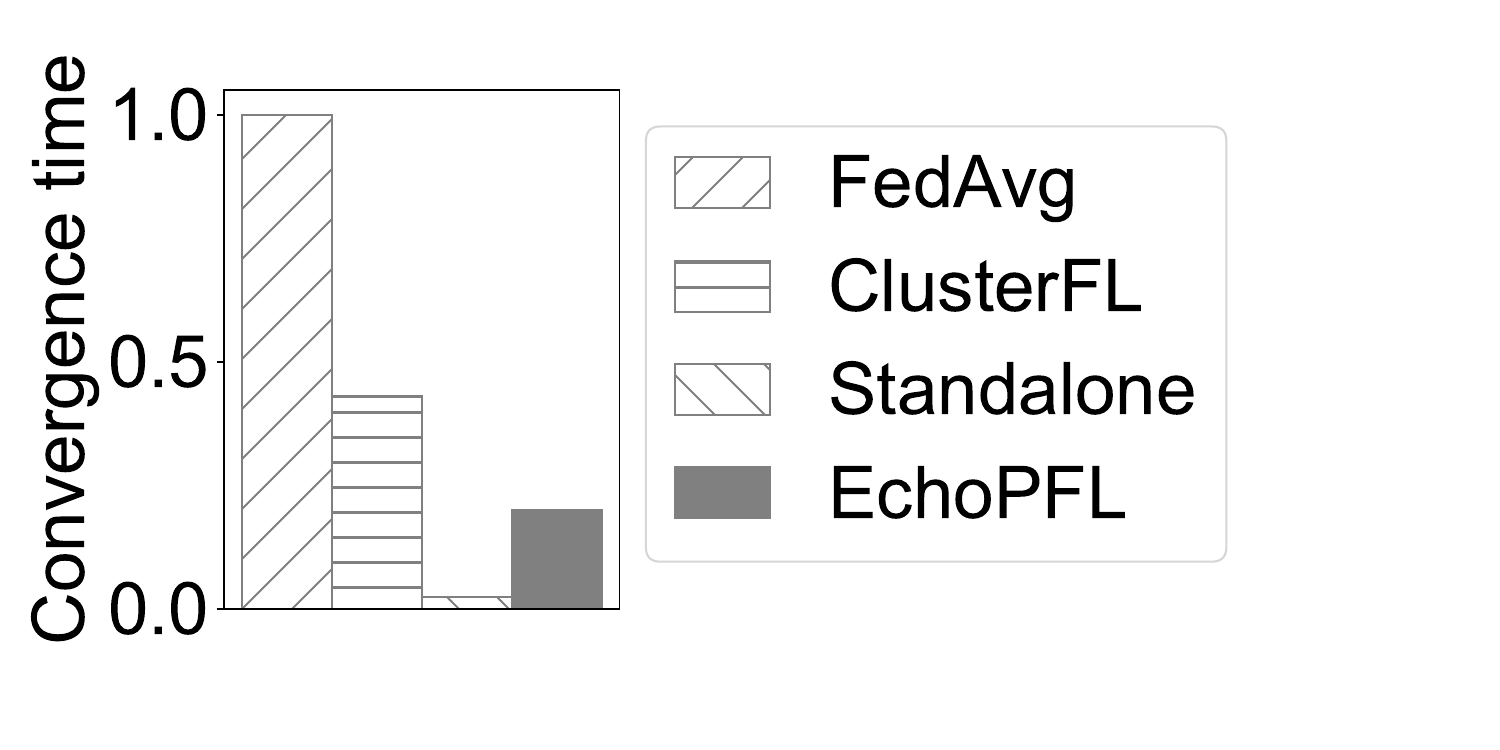}
    \vspace{-10pt}
    \label{fig:scenario_d_time}
  }
  \vspace{-10pt}
  \caption{Training time comparison in four scenarios.}
  \vspace{-10pt}
  \label{fig:scenario_time}
\end{figure*}

\subsubsection{Intermediate Clustering Result}
We use the synchronous clustering results of ClusterFL \cite{ouyang2021clusterfl} for comparison because ClusterFL can access all client weights for optimal clustering.
We test with task $T_1$.
First, as shown in \figref{fig:cluster_result}, the clusters identified by \sysname are similar to those of ClusterFL.
Their cosine similarity reaches up to $99\%$ (see \figref{fig:cluster_end_result}(a)), showcasing the efficacy of \sysname's dynamic clustering.
Second, we investigate the impact of the initial cluster number in \sysname on the resulting clusters in \figref{fig:cluster_end_result}(a). 
\sysname's is resilient to different initial cluster numbers.
Furthermore, as shown in \figref{fig:cluster_end_result}(b), the initial cluster number has a minimal impact ($0.5\%$) on training time and accuracy.


\subsection{Performance in Real-world Mobile Scenarios}
We tested \sysname and three baseline methods across four real-world mobile scenarios with $T_1$ task.
\tabref{exp:scenario_tab} outlines the configurations for data and device heterogeneity at each mobile client in the four scenarios, where diverse subsets of classes are assigned to different devices from $D_1$ to $D_3$.
In Scenario B, where data heterogeneity is the focus (\figref{fig:scenario_b}), \sysname's accuracy outperforms both FedAvg and the Standalone baseline due to personalization. 
As data heterogeneity increases from client 1 to 5, the accuracy advantage becomes more pronounced, surpassing ClusterFL at client 5.
In Scenario C, which highlights device heterogeneity (\figref{fig:scenario_c}), \sysname outperforms the baselines, mainly on fast devices (Client 3, 4, and 5). 
It indicates that \sysname enables fast devices to release highly accurate updated models promptly.
In Scenario D, with both device and data heterogeneity (\figref{fig:scenario_d}), \sysname achieves notably higher accuracy than the baselines. 
It is attributed to \sysname's personalization capability and its asynchronous mechanism.
Furthermore, across all these scenarios, \sysname consistently exhibits the shortest training time, as demonstrated in \figref{fig:scenario_time}.

\begin{figure}[t]
  \centering
  \subfloat[Accuracy]{
    \includegraphics[width=0.28\textwidth]{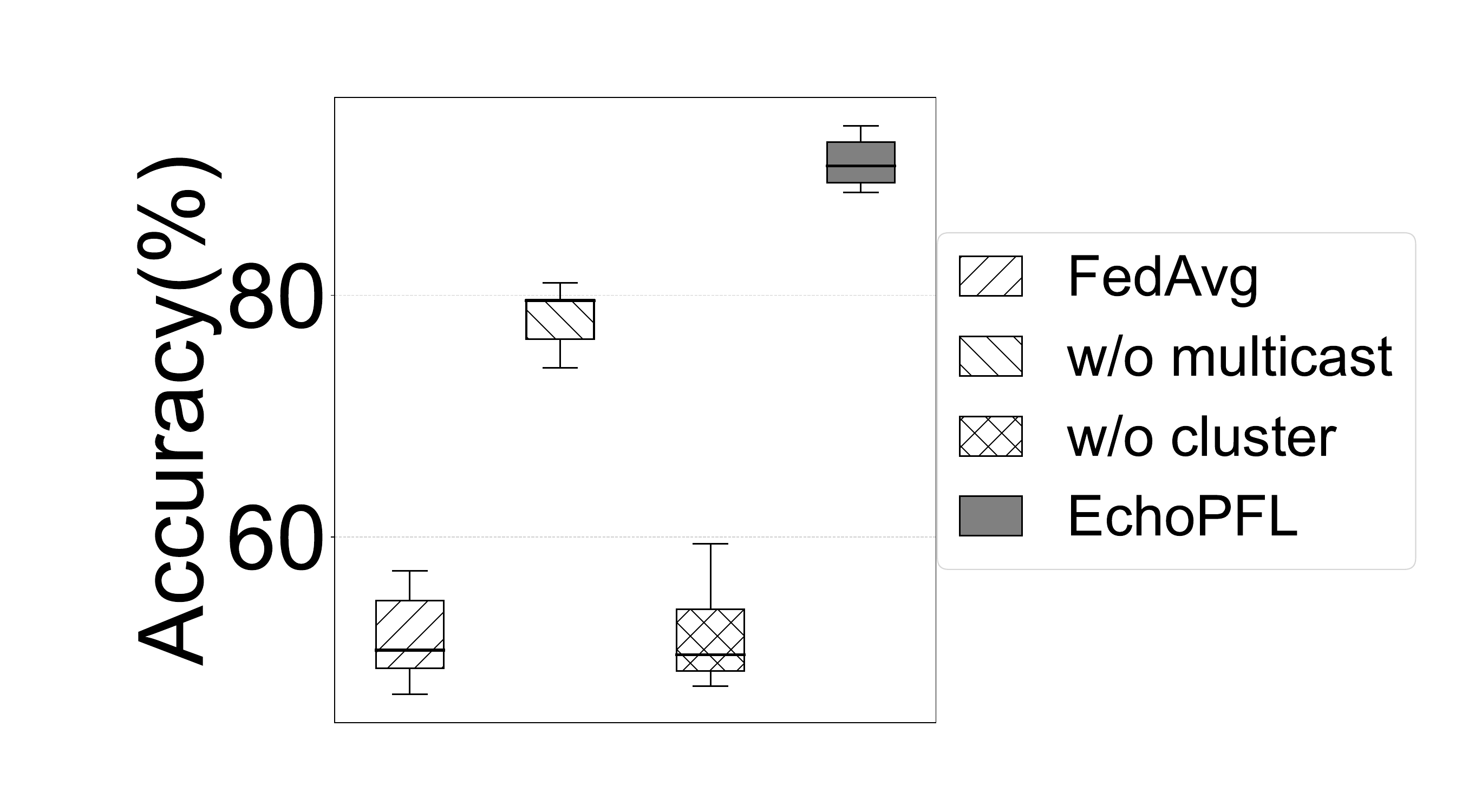}
    \vspace{-10pt}
    \label{fig:boardcast benchmark(a)}
  }
  \hspace{20mm}
  \subfloat[Training time]{
    \includegraphics[width=0.28\textwidth]{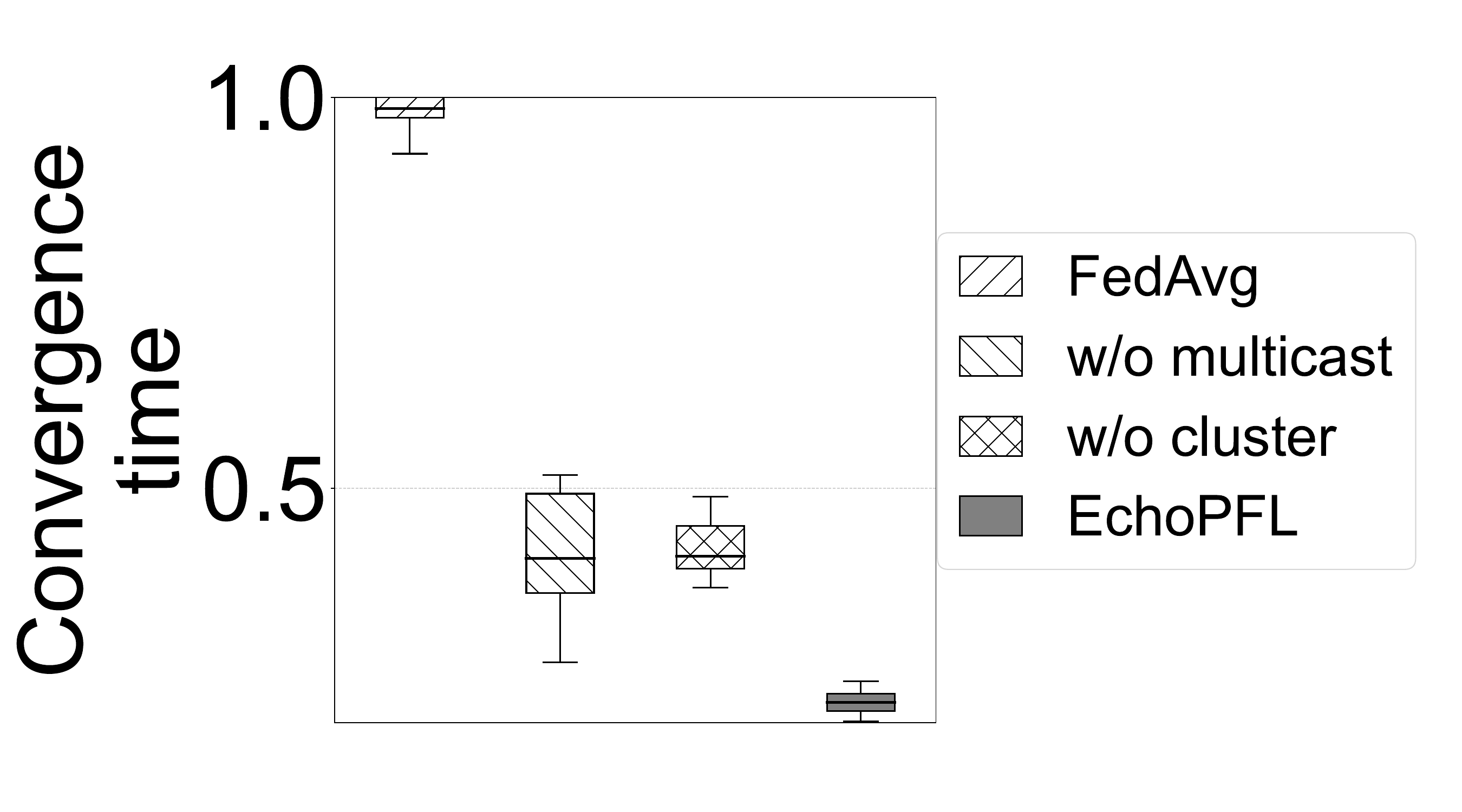}
    \vspace{-10pt}
    \label{fig:boardcast benchmark(b)}
  }
  \vspace{-10pt}
  \caption{The impact of the clustering, and broadcast mechanisms on EchoPFL's performance.}
  \vspace{-10pt}
  \label{fig:boardcast benchmark}
\end{figure}

\begin{table}[t]
\scriptsize
\caption{Time and accuracy differences Using L1 distance and KL divergence in real-time cluster partition}
\vspace{-10pt}
\begin{tabular}{|c|c|c|c|}
\hline
                                                                               & \textbf{\begin{tabular}[c]{@{}c@{}}Time for  each round (s)\end{tabular}} & \textbf{\begin{tabular}[c]{@{}c@{}}Trainig   time (min)\end{tabular}} & \textbf{\begin{tabular}[c]{@{}c@{}} Accuracy  (\%)\end{tabular}} \\ \hline
\textbf{\begin{tabular}[c]{@{}c@{}}L1-distance\\ in incremental clustering\end{tabular}}   & 0.0011                                                                    & \multirow{2}{*}{78.9}                                                    & \multirow{2}{*}{90.1}                                                   \\ \cline{1-2}
\textbf{\begin{tabular}[c]{@{}c@{}}KL divergence\\ in cluster adjustment\end{tabular}}     & 0.105                                                                     &                                                                          &                                                                         \\ \hline
\textbf{\begin{tabular}[c]{@{}c@{}}KL divergence\\ in incremental clustering\end{tabular}} & 13.2                                                                      & 421.2                                                                    & 90.4                                                                    \\ \hline
\end{tabular}
\label{exp:L1_KL}
\end{table}

\begin{figure}[t]
  \centering  \includegraphics[width=0.49\textwidth]{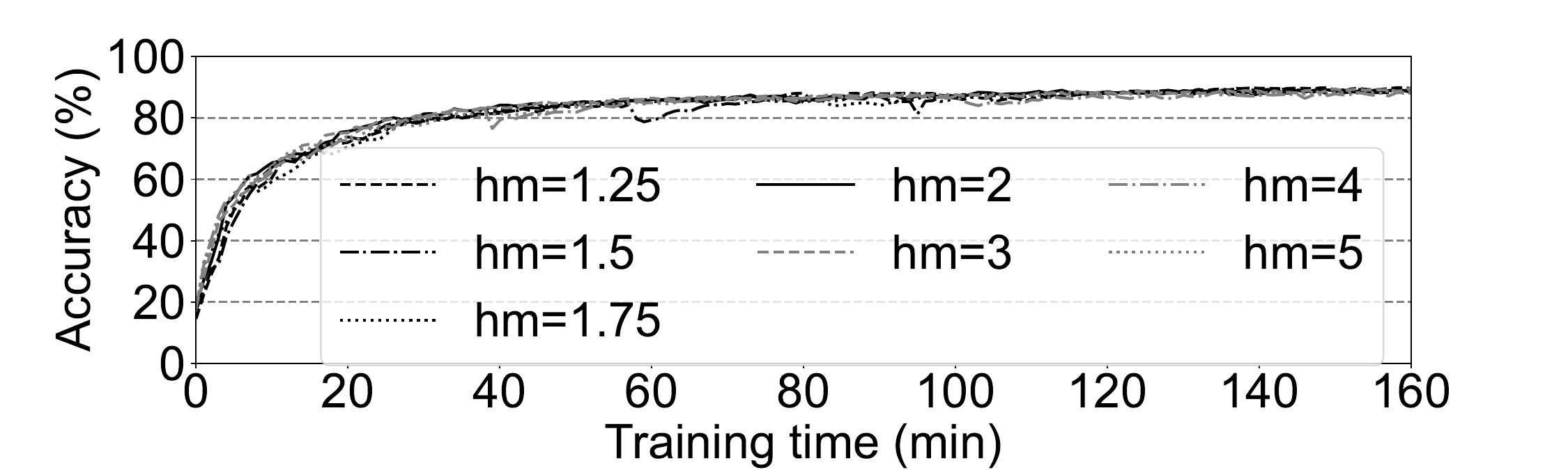}
  \vspace{-10pt}
  \caption{Impact of $hm$ in cluster merging operation.}
  \label{fig:hyperparameter_merge}
  \vspace{-10pt}
\end{figure}

\subsection{Ablation and Micro-benchmark}
\label{sec:benchmark}

This subsection validates \sysname's module and explores different hyperparameter settings. 

\subsubsection{w/ $vs.$ w/o clustering:} 
we assess the necessity and impact of the dynamic client clustering method in \sysname, which is responsible for customizing personalized models.
As depicted in \figref{fig:boardcast benchmark}, without the dynamic clustering block, the accuracy is reduced to be similar to FedAvg.

\subsubsection{w/ $vs.$ w/o broadcast}
We validate the significance of the in-cluster model broadcast method in \sysname.
As shown in \figref{fig:boardcast benchmark}, without the broadcast method, the accuracy decreased by 8.09\%, and the training time is prolonged by 1.8$\times$.

\subsubsection{Distance measure choice in real-time cluster partition}
We validate the choice of distance measures in client clustering.  
As shown in \tabref{exp:L1_KL}, using L1 distance results in decreases up to $5 \times$ training time compared to KL divergence, satisfying the real-time demands. 
And leveraging KL divergence for the merge step allows \sysname to achieve high final accuracy.

\subsubsection{Cluster number hyperparameter in merging} 
As discussed in \secref{sec:cluster_adjustment}, the maximized cluster number hyperparameter plays a crucial role in determining when should trigger the merge operation.
This experiment validates \sysname's robustness to diverse hyperparameters $hm \times$ for triggering the merge operation. 
\figref{fig:hyperparameter_merge} demonstrates that \sysname is robust and insensitive to the value of $hm$ in convergence time and accuracy. 
Thus we set $hm$ to 2 by default in \sysname. 

\subsection{Real-world Case Study}
\label{sec:case_study}

\begin{figure*}[t]
    \centering
    \includegraphics[width=0.6\textwidth]{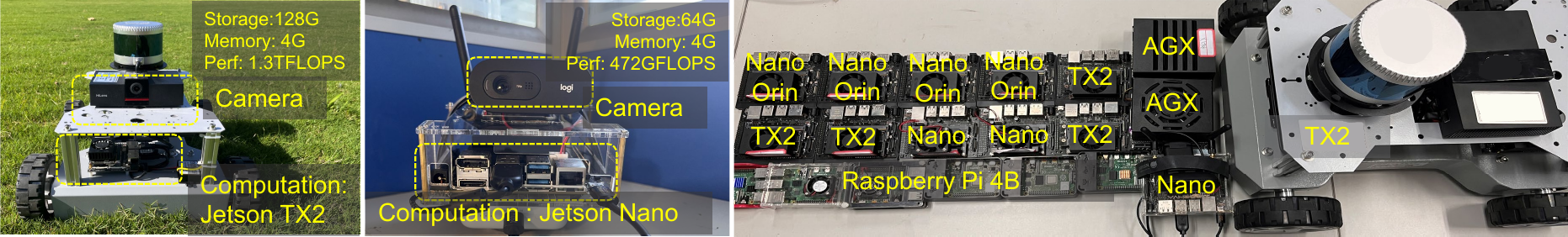}
    \caption{Illustration of the case study with 20 ubiquitous mobile devices.}
    \label{fig:case_all}
    \vspace{-12pt}
\end{figure*}

\begin{figure*}[t]
    \centering
    \includegraphics[width=0.63\textwidth]{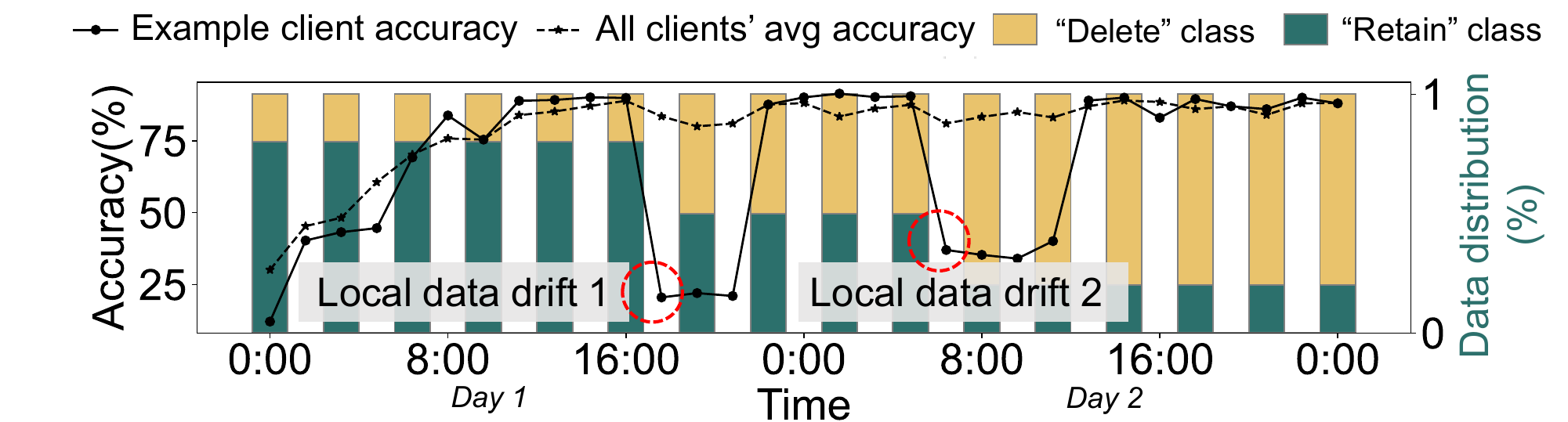}
    \caption{Average local testing accuracy across all mobile clients and local testing accuracy at an example client under Non-IID data with dynamic distribution shifts.
    }
    \vspace{-15pt}
    \label{fig:case_acc}
\end{figure*}

\begin{figure*}[t]
    \centering    
    \includegraphics[width=0.63\textwidth]{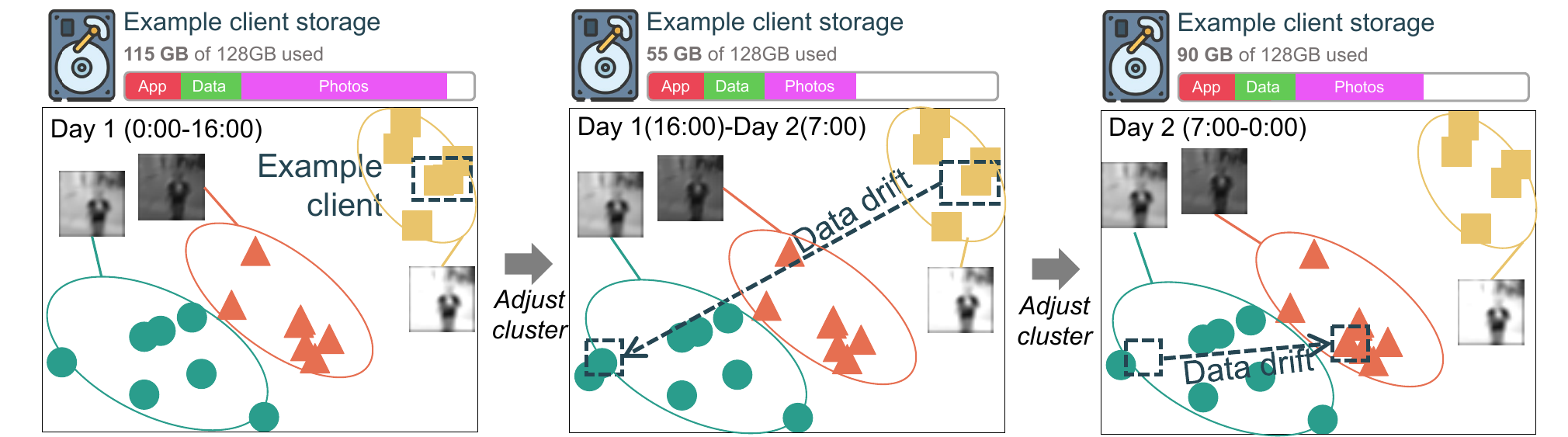}
    \vspace{-10pt}
    \caption{Visualizing dynamically adjusted clusters.
    }
    \label{fig:case_cluster}
    \vspace{-20pt}
\end{figure*}

\rev{We adopt 20 mobile and embedded devices($D_{1}$, 5$D_{2}$, 4$D_{3}$,2$D_{4}$ and 6$D_{5}$) to conduct a two-day study with the automatic image file cleaning applications ($T_4$), as shown in \figref{fig:case_all}.}
This app can help users automatically clean redundant images/videos in the image file library.
We employed 20 participants to label each image sample collected locally from 20 embedded devices as either ``Delete'' or ``Retain'' class based on their preferences, which resulted in Non-IID data distributions across devices.
The labeled data from each device is split into testing and training data by 2:8.
In addition, to validate the adaptability of \sysname, we directed participants to change their labeling preferences for the "retain/delete" classes twice during the 2-day study, thereby simulating local data distribution shifts. 

\figref{fig:case_acc} shows an example of client's data distribution shift on Day 1 (16:00) and Day 2 (7:00). 
In this two-day study, the example client's models dynamically adapted to the changing data distribution during federated learning.
The average local testing accuracy across all Non-IID clients remained stable and consistently exceeded 80\%, affirming \sysname's efficacy in managing diverse data.
Furthermore, during the shifts in local data distribution, the accuracy notably dropped at 16:00 on Day 1 and 7:00 on Day 2. 
Our approach promptly responded with adjustments over 2-3 rounds of federated training, allowing the personalized models to adapt to the new data. 
Consequently, the local testing accuracy swiftly rebounded to 89.3\%.
We also showcased the adaptability in \sysname's clusters in \figref{fig:case_cluster}. 
To visualize, we adopt principal component analysis to reveal how \sysname's clusters adapt to the changing data distribution of this example client.

%% file: body/related.tex
\section{Related Work}

\fakeparagraph{FL in Ubiquitous Applications}
Recently, federated learning (FL)~\cite{mcmahan2017communication} has been widely applied in mobile scenarios~\cite{pfeiffer2023federated}. 
Researchers deployed FL in various mobile scenarios, including transportation ~\cite{liang2022federated, he2022automatch, wang2022fed}, recommendation systems~\cite{niu2020billion},  activity recognition~\cite{ouyang2021clusterfl, ouyang2023harmony} and robotics \cite{liu2019lifelong}. 
For example, Niu \etal \cite{niu2020billion} employed FL to enhance the recommendation system within a mobile system operating at a billion-scale.
To ease development across diverse mobile applications, Beutel \etal \cite{beutel2020flower} proposed a system framework called Flower. 

\fakeparagraph{Personalized FL}
Due to Non-IID and unbalanced client data, personalized FL (PFL) is proposed \cite{tan2022towards} to output multiple personalized models instead of a single one. 
Specifically, PFL contains two main categories: global model personalization and learning personalized models. 
The former always involves “FL training + local adaptation” steps.
It mainly contains local fine-tuning \cite{zhang2022fine, wang2023fedftha, zeng2022gradient} and meta-learning \cite{fallah2020personalized}. 
For example, Wang \etal~\cite{wang2023fedftha} propose to fine-tune the head layers of the global model to realize personalization.
The latter introduces methods like clustering~\cite{ghosh2020efficient,briggs2020federated,ouyang2021clusterfl}, multi-task learning~\cite{smith2017federated}, and model interpolation~\cite{hanzely2020federated}.
\rev{
Additionally, \cite{jothimurugesan2023federated} introduced dynamic clustering to achieve personalization in the presence of various types of drift.  
\sysname focus on clustering-based PFL algorithms arises from the high clusterability observed in the data distributions across many mobile applications \cite{cao2018gchar, stisen2015smart, ouyang2021clusterfl}. 
Also, the cluster-based methods for handling device heterogeneity can be applied to handle these drifts as well.
}

\fakeparagraph{Synchronous and Asynchronous FL}
Device heterogeneity refers to the diversity in computational resources. 
It can be addressed by modifying the model architecture and system-level adaptation.
The former assigns lightweight models to devices with lower  resources~\cite{rapp2022distreal,diao2020heterofl,horvath2021fjord,li2019fedmd,tan2022fedproto}. 
For example, Li \etal \cite{li2019fedmd} utilize knowledge distillation to aggregate models.
The latter mainly fall into client selection-based methods~\cite{lai2021oort,li2022pyramidfl}, asynchronous FL (AFL)\cite{xie2019asynchronous,chen2020asynchronous,gu2021privacy}, and semi-asynchronous FL (SAFL) methods~\cite{wu2020safa,sun2022fedsea}.
PyramidFL~\cite{li2022pyramidfl} selects clients with similar performance to participate in the same training round.
\rev{\cite{xiang2023towards} present to adjust the broadcasting timing before or after local training in a synchronous setting, with a fixed broadcast frequency. }
For SAFL, Sun \etal~\cite{sun2022fedsea} propose FedSEA, establishing periodic synchronization points to mitigate the significant impact caused by stragglers. 
For AFL, Xie \etal~\cite{xie2019asynchronous} let the server immediately aggregate uploaded weights from each client while compromising the knowledge aggregation from slow devices.
\sysname embraces the asynchronous paradigm for its latency benefits and also addresses its shortcoming in aggregating data from stragglers by employing on-demand broadcast.

\section{Conclusion}
This paper presents \sysname, a client-server coordination mechanism for asynchronous personalized FL (PFL) via on-demand model broadcast.
\sysname is the first work that effectively integrates the asynchronous mechanism into PFL, ensuring the inclusion of all \rev{fast or slow mobile clients} without sacrificing accuracy.
It incrementally creates and manages clusters based on the incoming model updates and feedback. 
And it predicts the optimal broadcast frequency to further reduce the downstream communication cost without compromising accuracy.
Evaluations on four popular mobile tasks and real-world scenarios over twenty mobile devices show that \sysname achieves training time reductions of up to 88.2\%, accuracy improvements of up to 41.04\%, and communication cost reduction of up to 37\%. 
\rev{\sysname enables diverse ubiquitous devices to efficiently participate in the federated learning process, meeting personalization needs, and providing systematic support for the universal application of Federated Learning (FL) systems.}
\rev{In future work, we aim to improve \sysname's compatibility with various PFL algorithms, addressing feature and concept skew. Additionally, we intend to integrate \sysname into deep learning frameworks, facilitating the smooth deployment of asynchronous PFL in real-world ubiquitous applications.}
